\pdfoutput=1

\documentclass[11pt]{article}

\usepackage[preprint]{acl}

\usepackage{times}
\usepackage{latexsym}

\usepackage[T1]{fontenc}

\usepackage[utf8]{inputenc}

\usepackage{microtype}

\usepackage{inconsolata}

\usepackage{graphicx}

\usepackage{booktabs}
\usepackage{array}
\usepackage{multirow} 
\usepackage{makecell}

\title{\textsc{Primus}: A Pioneering Collection of Open-Source Datasets for Cybersecurity LLM Training}

\author{
  \normalsize\textbf{Yao-Ching Yu}\thanks{\scriptsize Primary Contributor.}, 
  \textbf{Tsun-Han Chiang}\footnotemark[1]\thanks{\scriptsize Equal Contribution.}, 
  \textbf{Cheng-Wei Tsai}\footnotemark[1]\footnotemark[2], 
  \textbf{Chien-Ming Huang}\footnotemark[1]\footnotemark[2], 
  \textbf{Wen-Kwang Tsao}
 \\
 \normalsize AI Lab, TrendMicro
 \\
 \normalsize{\texttt{\{yaoching\_yu,james\_chiang,dennis\_tsai,liam\_huang,spark\_tsao\}@trendmicro.com}}
 }
 
\begin{document}
\maketitle
\begin{abstract}
Large Language Models (LLMs) have shown remarkable advancements in specialized fields such as finance, law, and medicine. However, in cybersecurity, we have noticed a lack of open-source datasets, with a particular lack of high-quality cybersecurity pretraining corpora, even though much research indicates that LLMs acquire their knowledge during pretraining. To address this, we present a comprehensive suite of datasets covering all major training stages, including pretraining, instruction fine-tuning, and reasoning distillation with cybersecurity-specific self-reflection data. Extensive ablation studies demonstrate their effectiveness on public cybersecurity benchmarks. In particular, continued pre-training on our dataset yields a \emph{\textbf{15.9\%}} improvement  in the aggregate score, while reasoning distillation leads to a \emph{\textbf{15.8\%}} gain in security certification (CISSP). We will release all datasets and trained cybersecurity LLMs under the ODC-BY and MIT licenses to encourage further research in the community.\footnote{\scriptsize \textbf{For access to all datasets and model weights, please refer to this \href{https://huggingface.co/collections/trendmicro-ailab/primus-67b1fd27052b802b4af9d243}{link}.}}
\end{abstract}

\section{Introduction}
Large Language Models (LLMs) have significantly advanced artificial intelligence by leveraging massive data and sophisticated neural architectures, such as \emph{ChatGPT} \cite{ouyang2022training}, \emph{Llama} \cite{dubey2024llama} and \emph{DeepSeek} \cite{guo2025deepseek}. These models excel at understanding and generating human language \cite{wei2022emergent,minaee2024large} and adapt well when collaborating with domain experts \cite{ge2023openagi}, enabling tailored applications in fields like medicine, law, and education \cite{lai2024large,zhou2023survey,yan2024practical}. Meanwhile, in cybersecurity, as cyber threats continue to evolve \cite{li2021comprehensive,ghelani2022cyber}, traditional methods such as signature- and rule-based systems are struggling to keep up. Advances in AI, particularly through LLMs, therefore offer promising new avenues for enhancing cybersecurity \cite{ferrag2024generative}.

\begin{figure}[t]
  \centering
  \includegraphics[width=\columnwidth]{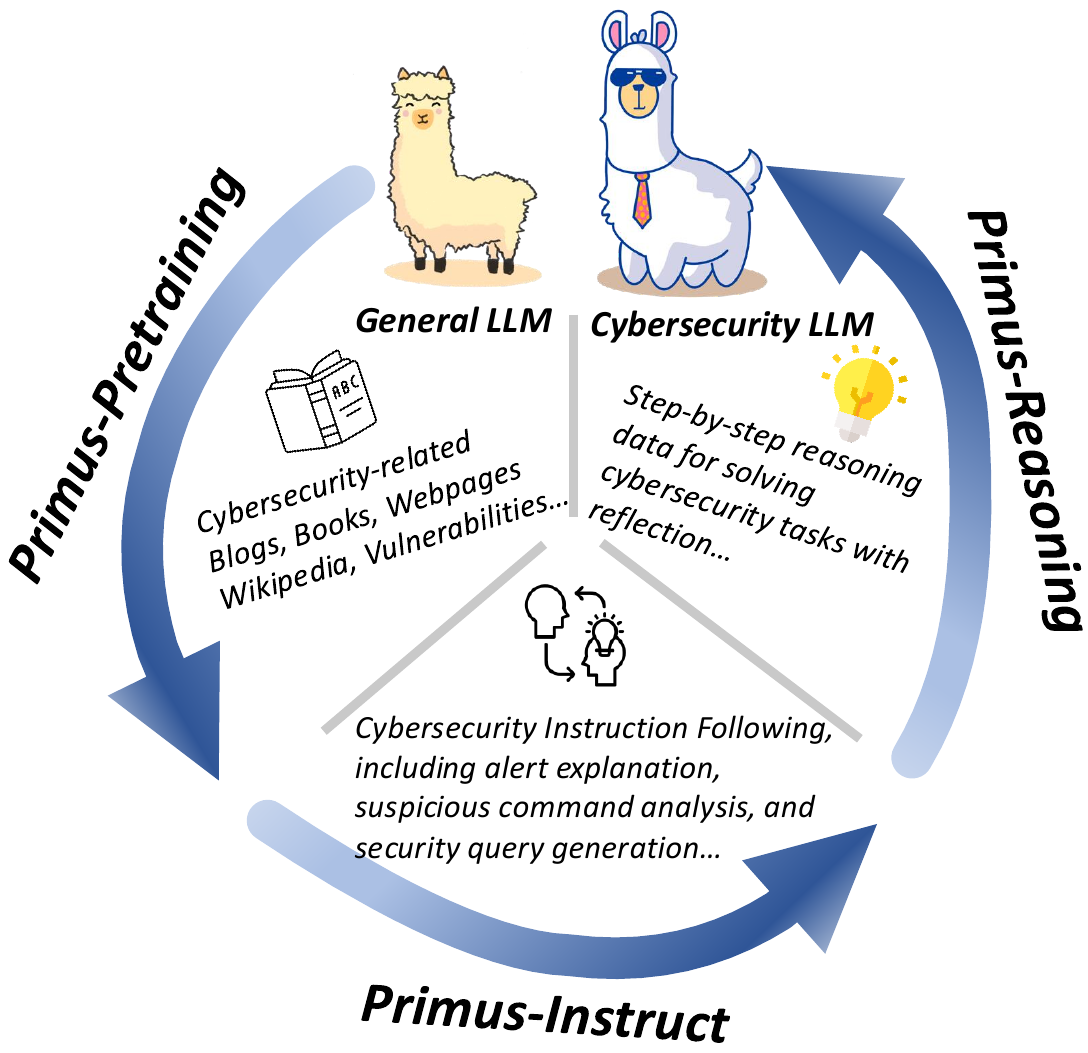}
  \caption{Overview of our training pipeline. \textsc{Primus-Pretraining}, \textsc{Primus-Instruct}, and \textsc{Primus-Reasoning} are the datasets of different training stages.}
  \label{fig:overview}
\end{figure}

Common training methods for LLMs include pre-training (PT) \cite{radford2018improving}, supervised fine-tuning (SFT) \cite{zhang2023instruction}, and reinforcement learning (RL) \cite{wang2024reinforcement}. Recent studies suggest LLMs acquire knowledge primarily during PT, and continued pre-training (CPT) \cite{gururangan-etal-2020-dont}, which further trains pre-trained models on large amounts of domain-specific text, can enhance their grasp of domain knowledge. In contrast, SFT may introduce hallucinations as new knowledge is learned \cite{gekhman-etal-2024-fine}. More recently, collecting reflection data from reasoning models for distillation has also become a trend \cite{huang2024o1}. Typically, obtaining a domain-specific LLM may require applying multiple training methods, as in our pipeline (Fig.\ref{fig:overview}).

\begin{figure}[t]
  \centering
  \includegraphics[width=\columnwidth]{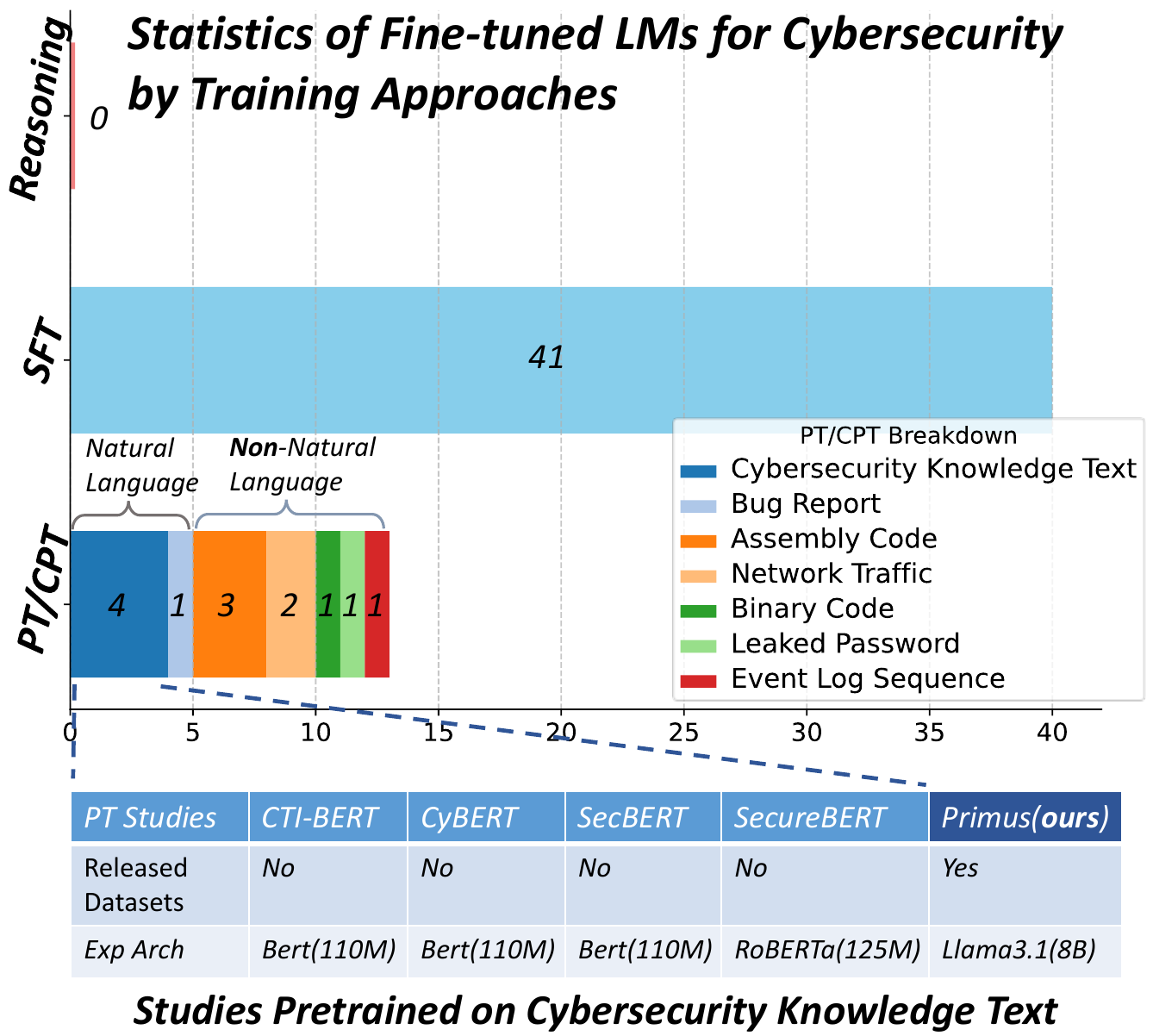}
  \caption{Motivation behind \textsc{Primus}. Statistics of existing cybersecurity language models, where \emph{reasoning} means training models to reason via distillation or RL.}
  \label{fig:stats_cyber_lm}
\end{figure}

The cybersecurity field has yet to fully benefit from this transformative technology, which requires domain expertise due to its broad and complex nature. Our statistics on cybersecurity LLM survey papers \cite{zhang2024llms,xu2024large} indicate that most existing research focuses on SFT to align model outputs, while PT or CPT is largely performed on non-natural language data such as assembly code \cite{jiang2023nova,wang2024clap,sun2023dexbert}, as shown in Fig.\ref{fig:stats_cyber_lm}. Clearly, these approaches have limited effectiveness in improving the general cybersecurity knowledge of LLMs. On the other hand, models pre-trained on cybersecurity knowledge \cite{park2023pretrained,ranade2021cybert,secbbert_github,aghaei2022securebert} are limited to small ones like BERT \cite{kenton2019bert}, and none of them have released datasets. To the best of our knowledge, LLMs pre-trained on cybersecurity knowledge or distilled on reasoning data from cybersecurity tasks remain \emph{unexplored}.

To address this gap, we extend prior work on domain-specific LLMs like medicine \cite{labrak-etal-2024-biomistral} and law \cite{colombo2024saullm} to cybersecurity. Our contributions are as follows:

\vspace{0.2\baselineskip}
\noindent\textbullet~\textbf{\emph{A Collection of Cybersecurity Datasets.}} We create a series of carefully curated datasets covering multiple stages of LLM training, including pre-training (\textsc{Primus-Pretraining}), instruction fine-tuning (\textsc{Primus-Instruct}), and reasoning fine-tuning (\textsc{Primus-Reasoning}), as shown in Fig.\ref{fig:overview}. Extensive ablation studies and evaluations on cybersecurity benchmarks show that these datasets can effectively improve cybersecurity capabilities. All\linebreak datasets will be released under the ODC-BY license to encourage further research in the community.

\vspace{0.2\baselineskip}
\noindent\textbullet~\textbf{\emph{A Family of Cybersecurity LLMs.}} We present a\linebreak family of cybersecurity LLMs designed to tackle domain-specific challenges, including \emph{Llama-Primus-Base}, a model further pre-trained with cybersecurity knowledge based on \emph{Llama-3.1-8B-Instruct}, achieving a \textbf{\emph{15.9\%}} improvement on aggregated cybersecurity benchmarks; \emph{Llama-Primus-Merged}, an instruction-tuned variant merged with \emph{Llama-3.1-8B-Instruct}, which \textbf{retains general instruction-following capability} while significantly improving cybersecurity performance; and \emph{Llama-Primus-Reasoning}, which is distilled from reasoning steps with reflection generated by a larger reasoning LLM on cybersecurity tasks, providing it long-thought capabilities and yielding a \textbf{\emph{15.8\%}} gain on security certification. Likewise, all models will be released under the MIT license.

\section{Training Datasets}
\subsection{Overview}
We build our dataset in multiple stages. First, we collect high-quality cybersecurity texts from reputable sources to form \textsc{Primus-Seed} (Sec.\ref{sec:primus-seed}), which is valuable but covers only a small fraction of cybersecurity content on the web. To extend it, we train a cybersecurity text classifier using \textsc{Primus-Seed} as positive samples and sampled data from FineWeb~\cite{penedo2024the}, a refined version of Common Crawl~\cite{commoncrawl}, as negative samples. This classifier filters cybersecurity-related content from FineWeb, producing \textsc{Primus-FineWeb} (Sec.\ref{sec:primus-fineweb}). By combining both datasets, we derive \textsc{Primus-Pretraining}. Next, we introduce \textsc{Primus-Instruct} (Sec.\ref{sec:primus-instruct}), which contains about 1k carefully curated cybersecurity tasks and general dialogues for instruction fine-tuning (IFT). Finally, \textsc{Primus-Reasoning} (Sec.\ref{sec:primus-reasoning}) provides reasoning steps generated by a stronger reasoning LLM on cybersecurity tasks for distillation.

\subsection{\textsc{Primus-Seed}}
\label{sec:primus-seed}
\subsubsection{Composition}

\begin{table}[t]
\footnotesize
\centering
\setlength{\tabcolsep}{2pt}
\begin{tabular}{>{\raggedright\arraybackslash}p{3.4cm}rrr} 
    \toprule
    \textbf{Category} & \textbf{Samples} & \textbf{Tokens} & \textbf{\textit{Avg.}} \\
    \midrule
    \multicolumn{4}{c}{\emph{Web Crawl / Official Dump}} \\
    \midrule
    Cybersecurity Blogs/News & 2,946 & 9,751,002 & 3,309.9 \\
    Cybersecurity Books & 6,499 & 2,910,464 & 447.8 \\
    Cybersecurity Companies Websites & 76,919 & 65,798,561 & 855.4 \\
    Cybersecurity Wikipedia & 6,636 & 9,567,196 & 1,441.7 \\
    MITRE & 3,432 & 2,435,118 & 709.5 \\
    \midrule
    \multicolumn{4}{c}{\emph{Expert Curation}} \\
    \midrule
    Campaigns & 136 & 37,106 & 272.8 \\
    Intrusion Sets & 343 & 60,524 & 176.5 \\
    Malware & 7,301 & 1,362,681 & 186.6 \\
    Reports & 11,317 & 934,954 & 82.6 \\
    Threat Actors & 27 & 2,264 & 83.9 \\
    Tools & 238 & 19,926 & 83.7 \\
    Vulnerabilities & 559,054 & 98,006,720 & 175.3 \\
    \midrule
    \textbf{Total} & 674,848 & 190,886,516 & 282.9 \\
    \bottomrule
\end{tabular}
\caption{Token statistics of different sources in the \textsc{Primus-Seed} dataset.}
\label{table:primus-seed}
\end{table}

We collect cybersecurity text through two main approaches. First, we gather data from reputable sources via official dumps or web crawling, converting raw HTML to readable Markdown using \texttt{dom-to-semantic-markdown}\footnote{\scriptsize \href{https://github.com/romansky/dom-to-semantic-markdown}{https://github.com/romansky/dom-to-semantic-markdown}}. Second, we incorporate curated cyber threat intelligence (CTI) manually collected by threat experts. The statistics of \textsc{Primus-Seed} are summarized in Tab.\ref{table:primus-seed}.

\paragraph{Official Dump and Web Crawl.}
We specifically collect cybersecurity-related text from diverse sources, including Blogs, News, Books, Websites, Wikipedia, and MITRE, guided by prior pretraining work~\cite{aghaei2022securebert}. For \textbf{Blogs} and \textbf{News}, we select content from government agencies, standards bodies, cybersecurity companies, media, and forums. Meanwhile, \textbf{Books} cover a wide range of cybersecurity topics, and we exclude covers, tables of contents, and appendices while treating each extracted page as a separate sample. We also collect \textbf{Webpages} from well-known cybersecurity companies, which may include product descriptions, company profiles, FAQs, and API documentation. In addition, \textbf{Wikipedia} does not provide a predefined cybersecurity subset, so we perform a custom filtering process. Each Wikipedia article is associated with one or more category tags, which can be further expanded into subcategory tags. Starting from the root category "\emph{Computer Security}", we recursively traverse its subcategories, using GPT-4o to determine whether a category is cybersecurity-related\footnote{\scriptsize The prompt is provided in the Appx.\ref{sec:appendix-prompts} (Fig.\ref{fig:prompt-wiki-category-classifier})}. This process yields 375 relevant categories, from which we extract corresponding Wikipedia articles. For \textbf{MITRE}, we leverage obsidian-mitre-attack\footnote{\scriptsize \href{https://github.com/vincenzocaputo/obsidian-mitre-attack}{https://github.com/vincenzocaputo/obsidian-mitre-attack}}, which converts STIX data from the official repository into readable Markdown.

\paragraph{Expert Curation.}
Another part of the data consists of CTI manually collected by our threat experts, categorized into Campaigns, Intrusion Sets, Malware, Threat Actors, Tools, Vulnerabilities, and Reports. Experts curate intelligence from open-source intelligence (OSINT), underground forums, and honeypots. OSINT includes public cybersecurity knowledge bases (e.g., MITRE ATT\&CK, CAPEC, CVE, CWE), government advisories (e.g., CISA, Europol), and threat intelligence sharing platforms that provide structured insight into attack patterns, vulnerabilities, and emerging threats. In addition, experts monitor underground forums for discussions of cybercriminal activity, while honeypots capture real-world attack data to enhance intelligence gathering.

\subsubsection{Preprocessing Pipeline}
Considering the varying quality of texts from different sources, we adopt a preprocessing pipeline inspired by previous dataset works~\cite{wenzek-etal-2020-ccnet,penedo2024the,raffel2019exploringc4}. Each source undergoes a dynamic combination of the following preprocessing steps.

\paragraph{LM Filtering.} 
We use perplexity from a language model trained on English Wikipedia as a quality score. Specifically, we use a 5-gram KenLM language model~\cite{heafield-2011-kenlm} due to its efficiency in processing large amounts of data. With this setup, we manually set an appropriate perplexity threshold for each source, and remove texts whose perplexity exceeds the threshold.

\paragraph{Deduplication.} 
Deduplication has been correlated with improvements in model performance~\cite{lee-etal-2022-deduplicating}. We adopt FineWeb's deduplication strategy, using a fuzzy hash-based approach with MinHash. Specifically, we extract 5-grams from each document and compute MinHashes using 112 hash functions, split into 14 buckets of 8 hashes each to target documents at least 75\% similar. Documents sharing the same 8 MinHashes in any bucket are considered duplicates.


\paragraph{C4 Filtering.} We also apply the quality filters from the C4 dataset \cite{raffel2019exploringc4}. Although being smaller than FineWeb, C4 performs well on certain benchmarks and remains a common component in the pretraining mix of recent models such as LLaMA1 \cite{touvron2023llama}. Its filtering rules include dropping lines without a terminal punctuation mark, mentioning javascript, or containing "\emph{terms-of-use}"/"\emph{cookie policy}" statements, and dropping documents that are too short or contain "\emph{lorem ipsum}" or a curly bracket (\texttt{\{}). We apply all of these filters except for the terminal punctuation and curly bracket filters.

\paragraph{Heuristic Filtering.}  
In addition to the above filters, we manually inspect each source and develop heuristic rules to further remove low-quality documents and outliers. For example, text containing phrases such as "\emph{Your download will begin in a few seconds}" will be dropped.

\subsubsection{Augmentation}  

We find that some web-scraped data contains valuable information but suffers from poor readability due to irregular formatting, such as inconsistent line breaks. To address this, we adopt a rewriting approach inspired by Cosmopedia\footnote{\scriptsize \href{https://github.com/huggingface/cosmopedia}{https://github.com/huggingface/cosmopedia}}, a reproduction of the high-quality synthetic dataset used in phi-1.5~\cite{li2023textbooks}. Specifically, we prompt an LLM to rewrite the given text into a specific style, including blog posts, textbooks, and Q\&A formats\footnote{\scriptsize The prompt is provided in the Appx.\ref{sec:appendix-prompts} (Fig.\ref{fig:prompt-pt-augmentation})}. To increase diversity, the rewriting LLM is randomly selected from GPT-4o, Llama-3.1-405B-Instruct, and DBRX \cite{Mosaic2024DBRX}.

\subsection{\textsc{Primus-FineWeb}}
\label{sec:primus-fineweb}

\subsubsection{Cybersecurity Classifier} 
Despite our efforts to collect as much cybersecurity text as possible in \textsc{Primus-Seed}, it likely covers only a small fraction of the cybersecurity-related content on the internet. To further expand our dataset, we train a binary classifier based on TinyBERT~\cite{jiao-etal-2020-tinybert} to distinguish cybersecurity-related text from non-cybersecurity text and apply it to FineWeb, a cleaned dataset derived from Common Crawl. Specifically, we use \textsc{Primus-Seed} as positive samples. Since cybersecurity text is only a small fraction of the web, we randomly take ten times as many samples from FineWeb and use them as negative samples to balance the dataset.

\begin{figure}[t]
  \centering
  \includegraphics[width=\columnwidth]{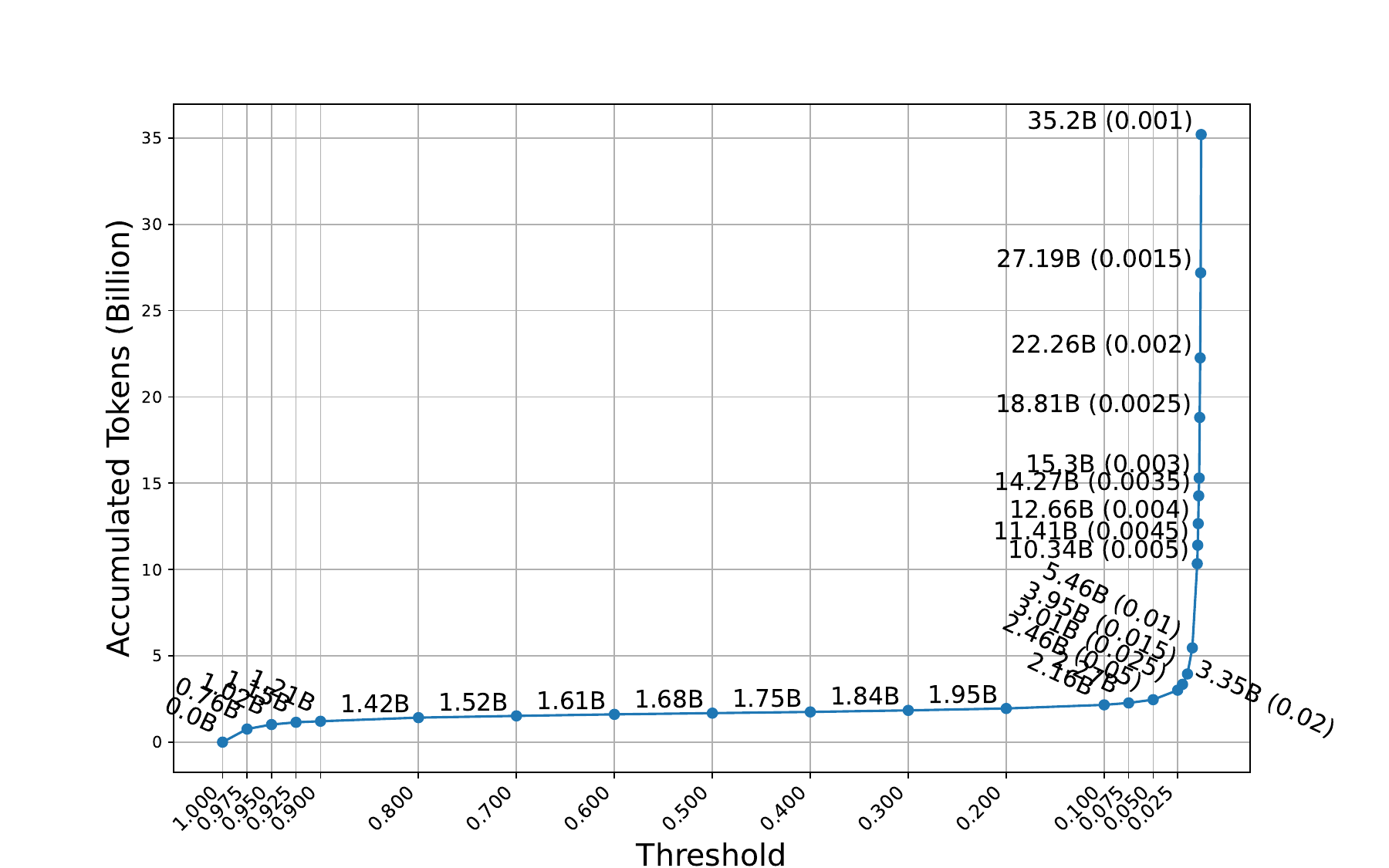}
  \caption{Cumulative token count in \textsc{FineWeb} for texts with a cybersecurity score exceeding various thresholds.}
  \label{fig:fineweb-threshold-tokens}
\end{figure}

\begin{figure}[t]
  \centering
  \includegraphics[width=\columnwidth]{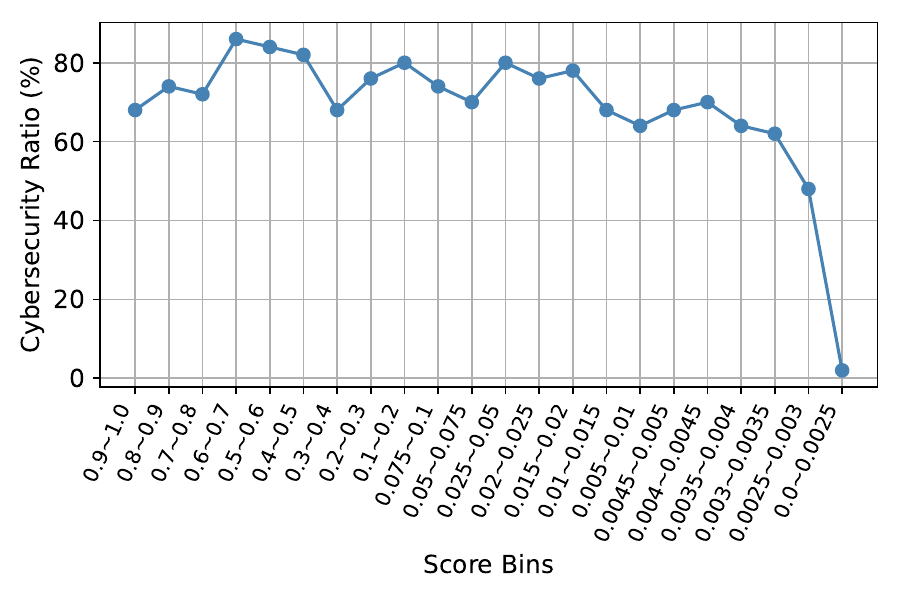}
  \caption{Ratio of cybersecurity-related text across different score bins in \textsc{FineWeb}.}
  \label{fig:fineweb-cyber-ratio}
\end{figure}

We then use the classifier to score all FineWeb texts on a scale from 0 to 1, where higher scores indicate greater cybersecurity relevance. The distribution in Fig.\ref{fig:fineweb-threshold-tokens} shows that lower scores correspond to a significant increase in text volume. To determine an appropriate threshold for filtering, we first verify that \textbf{\emph{whether texts with higher scores are truly cybersecurity-related}}. To do this, we leverage GPT-4o for accurate evaluation by dividing the scores into multiple bins, with dynamically adjusted bin sizes---smaller bins for lower scores---to account for the increased volume of data in lower score ranges. We randomly sample 50 texts from each bin and prompt GPT-4o\footnote{\scriptsize The prompt is provided in the Appx.\ref{sec:appendix-prompts} (Fig.\ref{fig:prompt-fineweb-cybersecurity-classifier})} for classification. As shown in Fig.\ref{fig:fineweb-cyber-ratio}, relevant text proportions remain above 60\% at higher scores, but drop below 50\% when scores fall below 0.003. Although incorporating some general text can help mitigate catastrophic forgetting~\cite{Sun2019LAMOLLM}, we prioritize maintaining a majority of cybersecurity content. Therefore, we set the final threshold at 0.003, which corresponds to 15.3B of FineWeb data.

\subsubsection{Deduplication Analysis}

\begin{table}[t]
\footnotesize
\centering
\setlength{\tabcolsep}{5pt}
\begin{tabular}{ccrrr} 
    \toprule
    \textbf{Threshold} & \textbf{\textit{Dedup.}} & \textbf{Samples} & \textbf{Tokens}& \textbf{\textit{Avg.}} \\
    \midrule
    0.003 & \emph{False} & 20,345,616 & 15.30B & 751.88 \\
    0.003 & \emph{True}  & 3,386,733  & 2.57B  & 759.11 \\
    0.9   & \emph{False} & 2,017,959  & 1.21B  & 600.37 \\
    0.9   & \emph{True}  & 393,154    & 0.23B    & 584.75 \\
    \bottomrule
\end{tabular}
\caption{Statistics of token counts before and after deduplication at different thresholds in the FineWeb.}
\label{table:primus-fineweb}
\end{table}

\begin{figure}[t]
  \centering
  \includegraphics[width=\columnwidth]{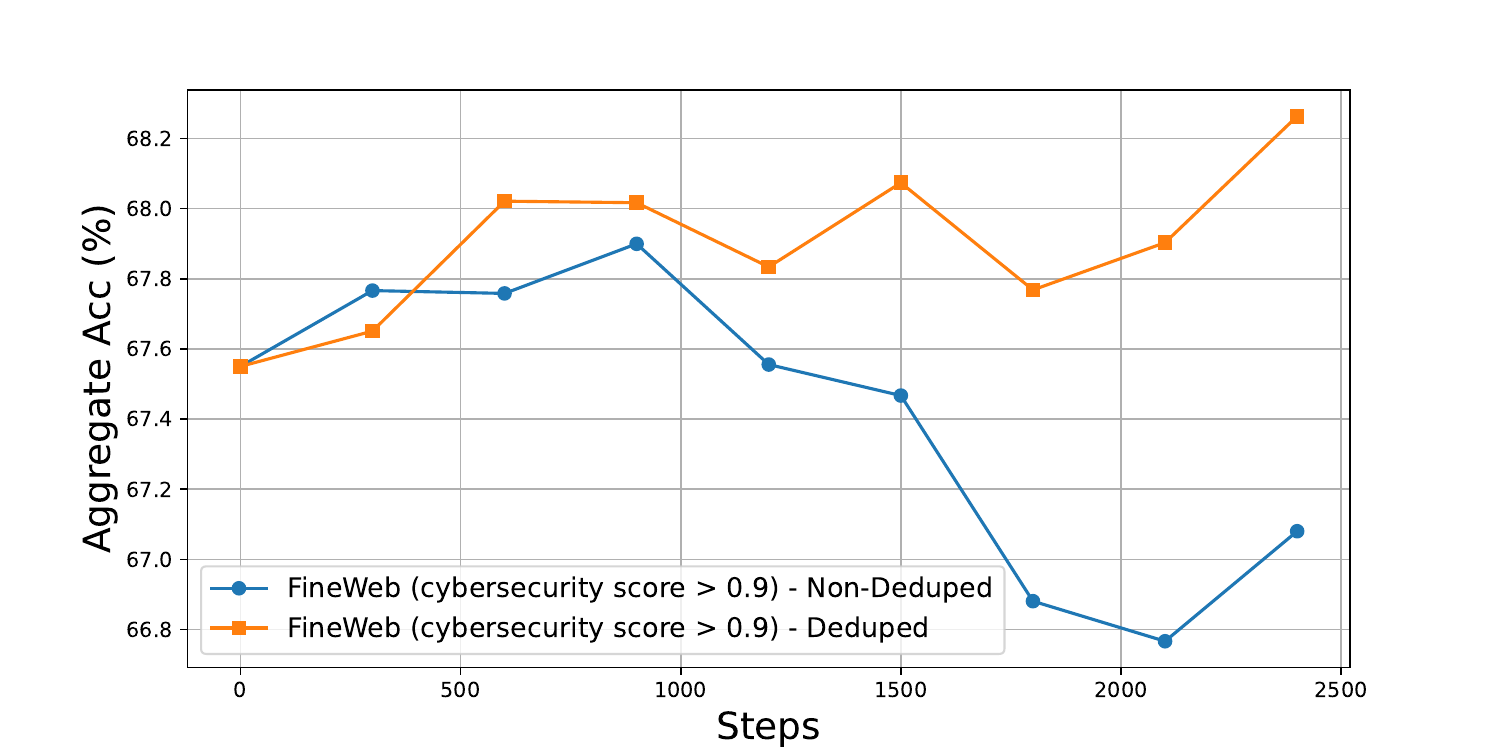}
  \caption{Comparison of deduplication on FineWeb cybersecurity data filtered at a classifier threshold 0.9.}
  \label{fig:fineweb-dedup-ablation}
\end{figure}

Upon inspecting the 15.3B dataset, we observed a significant amount of duplicate content. This occurs because FineWeb's ablation study found that deduplicating each Common Crawl snapshot separately yields better results than global deduplication, so FineWeb does not apply global deduplication. However, since our filtered dataset is much smaller, we conducted our own ablation study. Specifically, we extracted and deduplicated 1.21B tokens with a score above 0.9, reducing the number to 0.23B (pre- and post-deduplication token counts are listed in Tab.\ref{table:primus-fineweb}), and we also sampled 0.23B tokens directly from the 1.21B set as an undeduplicated control group. We pre-trained Llama-3.1-8B-Instruct for two epochs on both datasets and found that the deduplicated dataset significantly outperformed the undeduplicated one on our aggregate of multiple-choice question (MCQ) cybersecurity tasks (to be introduced in Sec.\ref{sec:benchmarks}), as shown in Fig.\ref{fig:fineweb-dedup-ablation}. Based on this observation, we finalized \textsc{Primus-FineWeb} with 2.57B deduplicated tokens filtered at a threshold of 0.003.

\subsection{\textsc{Primus-Instruct}}
\label{sec:primus-instruct}

\begin{table}[t]
\footnotesize
\centering
\setlength{\tabcolsep}{3pt}
\begin{tabular}{>{\raggedright\arraybackslash}p{5.5cm}r} 
    \toprule
    \textbf{Task} & \textbf{Samples} \\
    \midrule
    \multicolumn{2}{c}{\emph{Cybersecurity-related Tasks}} \\
    \midrule
    Alert Explanation & 100 \\
    Retrieved Security Doc QA & 100 \\
    Suspicious Command Analysis & 100 \\
    Security Event Query Generation & 100 \\
    Terraform Security Misconfiguration Fix & 96 \\
    \midrule
    \multicolumn{2}{c}{\emph{General (Multi-turn)}} \\
    \midrule
    General Instruction Following & 339 \\
    \bottomrule
\end{tabular}
\caption{Task distribution and corresponding sample counts in the \textsc{Primus-Instruct} dataset.}
\label{table:primus-instruct}
\end{table}

After pre-training, we use \textsc{Primus-Instruct} for instruction fine-tuning to restore the instruction-following capability of the model. To achieve this, we design several hundred cybersecurity tasks covering common business scenarios, including explaining detected alerts, answering questions about retrieved security documents, analyzing executed suspicious commands, generating query languages for retrieving security events, and providing security recommendations and risk assessments for Terraform configurations. Each example is answered by GPT-4o, and we further use Claude 3.5 Sonnet \cite{anthropic2024claude35sonnet} as a judge\footnote{\scriptsize The judge prompt is provided in the Appx.\ref{sec:appendix-prompts} (Fig.\ref{fig:prompt-primus-instruct-judge})} to discard samples with insufficiently helpful answers. In addition, we include several hundred multi-turn conversations on general topics generated by GPT-4o. As a result, these form \textsc{Primus-Instruct}, with statistics in Tab.\ref{table:primus-instruct}.

\subsection{\textsc{Primus-Reasoning}}
\label{sec:primus-reasoning}

With the release of OpenAI's reasoning model o1, an increasing number of studies have attempted to replicate its reasoning capabilities. One widely recognized approach is distillation, where reasoning samples with \emph{self-reflection} from existing reasoning models are used to guide models in acquiring long-thought capabilities~\cite{huang2024o1,liu2024deepseek}. To this end, we select cybersecurity reasoning tasks from CTI-Bench\footnote{\scriptsize A brief introduction to CTI-Bench is provided in Appx.\ref{sec:appendix-cti-bench}}~\cite{Alam2024CTIBench} and prompt o1-preview one to two times per question to generate solutions with reasoning steps and reflection\footnote{\scriptsize The prompt is provided in the Appx.\ref{sec:appendix-prompts} (Fig.\ref{fig:prompt-o1-reasoning})}, applying rejection sampling to retain only the correctly answered samples. We also include DeepSeek-R1, obtained by directly querying its open-source model to access reasoning steps. The dataset statistics are shown in Tab.\ref{table:primus-reasoning}.


\begin{table}[t]
\footnotesize
\centering
\setlength{\tabcolsep}{2pt}
\begin{tabular}{
    >{\raggedright\arraybackslash}p{2.2cm}   
    >{\centering\arraybackslash}p{1.2cm}     
    >{\centering\arraybackslash}p{1.9cm}     
    >{\centering\arraybackslash}p{1.7cm}     
}
\toprule
\multirow{2}{*}{\textbf{Dataset}}
  & \multirow{2}{*}{\textbf{Samples}}
  & \textbf{Accepted}
  & \textbf{\textit{Avg.\ Tokens}} \\
\cmidrule(lr){3-4}
  &  & \multicolumn{2}{c}{\makecell{(o1‐preview / DeepSeek‐R1)}} \\
\midrule
CTI-MCQ        & 1000 & 806 / 768   & 692 / 672   \\
CTI-RCM        & 1000 & 728 / 721   & 761 / 530   \\
CTI-RCM-2021   & 1000 & 635 / 683   & 766 / 543   \\
CTI-VSP        & 1000 & 231 / 312   & 1156 / 1395 \\
CTI-ATE        &   60 &   2 /   5   & 1314 / 1731 \\
\bottomrule
\end{tabular}
\caption{Statistics of the \textsc{Primus-Reasoning} dataset, distilled from o1-preview and DeepSeek-R1 on CTI-Bench questions, with only accepted correct samples.}
\label{table:primus-reasoning}
\end{table}

\section{Evaluation Protocol}  

This section introduces the cybersecurity benchmarks (Sec.\ref{sec:benchmarks}) and evaluation settings (Sec.\ref{sec:eval_settings}) used to assess training performance.

\subsection{Benchmarks}  
\label{sec:benchmarks}
To assess the performance and training effectiveness of \textsc{Primus} models, we evaluate them against seven cybersecurity benchmarks to measure their robustness and comprehensive understanding of security concepts, which we describe below.


\paragraph{CISSP.} The Certified Information Systems Security Professional (CISSP) is a widely recognized cybersecurity certification that assesses both technical expertise and managerial competence. We construct an evaluation set based on multiple-choice questions from CISSP learning materials.

\paragraph{CTI-Bench.} CTI-Bench is a benchmark for evaluating the reasoning and knowledge capabilities of LLMs in CTI. It consists of several subtasks, including CTI-RCM, CTI-VSP, CTI-ATE, and CTI-MCQ, which assess a model's ability to analyze vulnerabilities, infer security risks, extract attack techniques, and understand cybersecurity concepts.


\paragraph{CyberMetric.} CyberMetric~\cite{cybermetric} is a benchmark of human-verified multiple-choice questions designed to assess LLMs' cybersecurity knowledge across domains such as cryptography, network security, penetration testing, and compliance. We select a 500-question subset for evaluation as it is balanced and representative.

\paragraph{SecEval.} SecEval~\cite{li2023seceval} is a benchmark consisting of over 2,000 multiple-choice questions covering nine cybersecurity domains, including software security, cryptography, and network security. Generated by prompting GPT-4 with authoritative sources such as textbooks and official documentation, it provides a reliable measure of LLMs' cybersecurity proficiency.


\subsection{Evaluation Settings} 
\label{sec:eval_settings}

We integrate the above benchmarks into the \texttt{lm-evaluation-harness}~\cite{eval-harness} to ensure a standardized evaluation process. All evaluations are performed in the same environment to ensure fairness. We adopt the following two evaluation settings to evaluate models at different stages.

\paragraph{\emph{5-shot, w/o Chain-of-Thought (CoT).}} We prepend the first five questions from the benchmark along with their answers as context before the current question, guiding the model to output the correct answer directly instead of generating free-form responses. This setting is used to evaluate models after pretraining, when output formatting is more difficult to control.

\paragraph{\emph{0-shot, w/ CoT}.} We follow the evaluation setup from the OpenAI technical report benchmarks with \texttt{simple-eval}\footnote{\scriptsize \href{https://github.com/openai/simple-evals}{https://github.com/openai/simple-evals}}, using a standardized prompt\footnote{\scriptsize The prompt is provided in the Appx.\ref{sec:appendix-prompts} (Fig.\ref{fig:prompt-simple-evals})} that allows the model to articulate its reasoning before producing the final answer. Due to the formatting variability of CoT responses, we use GPT-4o-mini to extract the final answers before scoring.

\section{Training and Results}
\subsection{Overview}  
\label{sec:train-overview}

In this section, we present the entire training pipeline, which consists of four key stages. First, we expand the model's cybersecurity expertise and understanding through continued pre-training (Sec.\ref{sec:train-pretraining}), which reinforces key cybersecurity concepts and enables the model to provide accurate information on security threats and mitigation strategies. Next, we restore its instruction-following capability through instruction fine-tuning (Sec.\ref{sec:train-instruct-merge}), and further refine it through model merging to balance instruction-following and cybersecurity expertise. Finally, we train the model to develop reasoning capabilities on cybersecurity tasks (Sec.\ref{sec:train-reason-ft})\footnote{\scriptsize The training hyperparameters for each stage are provided in the Appx.\ref{sec:appendix-hyperparameters}}.

\subsection{Pre-Training}
\label{sec:train-pretraining}

\begin{table*}[ht]
\footnotesize
\centering
\setlength{\tabcolsep}{1.7pt} 
\begin{tabular}{>{\raggedright\arraybackslash}p{3.9cm} cccccccc} 
    \toprule
    \textbf{Model} & \textbf{CISSP} & \textbf{CTI-MCQ} & \textbf{CTI-RCM} & \textbf{CTI-VSP} & \textbf{CTI-ATE} & \textbf{CyberMetric} & \textbf{SecEval} & \textbf{\textit{Agg.}} \\
    \midrule
    Llama-3.1-8B-Instruct & 0.7073 & 0.6420 & 0.5910 & 1.2712 & 0.2721 & 0.8560 & 0.4966 & 2.29\\
    \hspace{0.5mm} \textsc{+ Primus-Seed} & 0.7132 & 0.6608 & 0.6100 & 1.2848 & 0.2829 & 0.8600 & 0.4998 & 2.34$\uparrow$2.1\% \\
    \hspace{0.5mm} \textsc{+ Primus-FineWeb} & 0.7191 & 0.6600 & 0.6680 & 1.1499 & 0.3006 & 0.8620 & 0.4984 & 2.56$\uparrow$11.5\% \\
    \hspace{0.5mm} \textsc{+ Primus-Seed+FineWeb} & \textbf{0.7230} & \textbf{0.6676} & \textbf{0.6780} & \textbf{1.0912} & \textbf{0.3140} & \textbf{0.8660} & \textbf{0.5007} & \textbf{2.66$\uparrow$15.9\%} \\
    \bottomrule
\end{tabular}

\caption{Performance of continued pretraining on Llama across cybersecurity benchmarks. The last three rows indicate pretraining with \textsc{Primus-Seed}, \textsc{Primus-FineWeb}, and their combination. CTI-VSP is scored using Mean Absolute Deviation \textbf{\emph{(lower is better)}}, CTI-ATE uses F1 score, and the others use accuracy. The aggregate score \emph{(Agg.)} is the sum of all benchmarks, with CTI-VSP negated. The best results are highlighted in \textbf{bold}.}

\label{table:primus-pretraining-performance}
\end{table*}

We use Llama-3.1-8B-Instruct as our base model due to its wide community adoption and strong performance at the same parameter scale. We perform continued pre-training on two cybersecurity datasets: \textsc{Primus-Seed} (Sec.\ref{sec:primus-seed}), which consists of curated cybersecurity text, and \textsc{Primus-FineWeb} (Sec.\ref{sec:primus-fineweb}), a filtered subset of cybersecurity content from FineWeb, to expand the model's cybersecurity expertise and understanding. To assess performance improvements, we evaluate the model against the seven cybersecurity benchmarks described in Sec.\ref{sec:benchmarks} (5-shot, w/o CoT).

We train the model using the \texttt{NeMo} \cite{NVIDIA_NeMo} on four 8$\times$H200 nodes, with training hyperparameters and details provided in Appx.\ref{sec:appendix-hyperparameters}. To analyze the impact of different datasets, we conduct an ablation study by pre-training the model separately on each dataset and jointly on both for two epochs. The results in Tab.\ref{table:primus-pretraining-performance} show that pre-training on either dataset improves the cybersecurity performance in the aggregate evaluation score. However, the largest improvement, \textbf{\emph{15.9\%}}, is observed when pre-training on the combined dataset, so we adopt this model as the Llama-Primus-Base for subsequent training stages\footnote{\scriptsize We also
 experimented with a 70B model in \textbf{Q2} of Appx.\ref{sec:appendix-faqs} (FAQs)}.

\subsection{Instruction Fine-Tuning and Merge}
\label{sec:train-instruct-merge}

\begin{table*}[t]
\footnotesize
\centering
\setlength{\tabcolsep}{1.2pt} 
\begin{tabular}{>{\raggedright\arraybackslash}p{3cm} ccccccccc} 
    \toprule
    \textbf{Model} & \textbf{CISSP} & \textbf{CTI-MCQ} & \textbf{CTI-RCM} & \textbf{CTI-VSP} & \textbf{CTI-ATE} & \textbf{CyberMetric} & \textbf{SecEval} & \textbf{MT-Bench} & \textbf{\textit{Agg.}} \\
    \midrule
    Llama-3.1-8B-Instruct & 0.7073 & 0.6420 & 0.5910 & 1.2712 & 0.2721 & 0.8560 & 0.4966 & \textbf{8.3491} & 4.11 \\
    Llama-Primus-Instruct & 0.7132 & \textbf{0.6660} & \textbf{0.6660} & \textbf{1.1161} & 0.3348 & 0.8640 & 0.4943 & 7.9063 & 4.21$\uparrow$2.4\% \\
    Llama-Primus-Merged & \textbf{0.7191} & 0.6656 & 0.6620 & 1.1233 & \textbf{0.3387} & \textbf{0.8660} & \textbf{0.5062} & 8.2938 & \textbf{4.33$\uparrow$5.4\%} \\
    \bottomrule
\end{tabular}
\caption{Performance comparison of Llama, the instruction-tuned Primus model, and their merge on cybersecurity and general benchmarks. The aggregated score \emph{(Agg.)} is computed as $0.3 \times$ MT-Bench + $0.7 \times$ aggregated cybersecurity score (sum of all benchmarks except MT-Bench, with CTI-VSP negated due to the use of Mean Absolute Deviation, where lower is better). The best results are highlighted in \textbf{bold}.}
\label{table:primus-instruction-performance}
\end{table*}

While Llama-Primus-Base gains enhanced cybersecurity knowledge and understanding from pre-training, it tends to perform text completion rather than follow instructions. To address this, we further fine-tune it using the \texttt{LLaMA-Factory} \cite{zheng2024llamafactory} on 4$\times$A100 GPUs for two epochs with \textsc{Primus-Instruct} (Sec.\ref{sec:primus-instruct}), a carefully curated mixed dataset of cybersecurity tasks and general conversations, resulting in Llama-Primus-Instruct. In addition to the cybersecurity benchmarks, we also introduce MT-Bench~\cite{zheng2023judging}, a multi-turn instruction-following evaluation benchmark spanning multiple domains using GPT-4 as a judge, which scores helpfulness on a scale of 1 to 10, allowing us to evaluate the overall instruction-following performance of the model. The results are shown in Tab.\ref{table:primus-instruction-performance}, where the MT-Bench score and the aggregated cybersecurity benchmark score are further aggregated with a weight of 30/70 in the rightmost column.

Llama-Primus-Instruct maintains its advantage in cybersecurity while achieving an MT-Bench score of 7.91. However, this remains lower than the 8.35 of Llama, resulting in a limited improvement in the aggregated score (2.4\%). To mitigate this, we apply DARE-TIES~\cite{yu2024languageDARES, yadav2023tiesmerging}, a model merging technique that balances diverse capabilities---specifically, instruction-following and cybersecurity expertise in our case. We conduct a grid search over the merging ratio, setting Llama-Primus-Instruct:Llama-3.1-8B-Instruct to $(0.5+w)$:$(0.5-w)$ and varying $w$ from 0 to 0.5 in steps of 0.05. The optimal ratio that maximizes the aggregated score is found to be 0.75:0.25, with the merged model chosen as Llama-Primus-Merged. Notably, this configuration retains cybersecurity performance comparable to Llama-Primus-Instruct while restoring the MT-Bench to 8.29, almost equal to Llama, resulting in a \emph{\textbf{5.4\%}} improvement in the aggregated score\footnote{\scriptsize We provide more details in \textbf{Q4} and \textbf{Q5} of Appx.\ref{sec:appendix-faqs} (FAQs)}.

\subsection{Reasoning Fine-Tuning}
\label{sec:train-reason-ft}

\begin{table}[t]
\footnotesize
\centering
\setlength{\tabcolsep}{0.5pt}           
\begin{tabular}{>{\raggedright\arraybackslash}p{4.05cm}
                >{\centering\arraybackslash}p{1.95cm}
                >{\centering\arraybackslash}p{1.6cm}}
    \toprule
    \textbf{Model} & \textbf{CISSP} & \textbf{\textit{Avg. Tokens}} \\
    \midrule
    \multicolumn{3}{c}{\emph{w/o CoT, 5-shot}} \\
    \midrule
    Llama-3.1-8B-Instruct           & 0.7073 & 1 \\
    Llama-Primus-Merged             & 0.7191 $\uparrow$1.67\% & 1 \\
    \midrule
    \multicolumn{3}{c}{\emph{w/ CoT, 0-shot}} \\
    \midrule
    Llama-3.1-8B-Instruct                           & 0.7288 $\uparrow$3.03\% & 279.69 \\
    \: + \textit{Distilled from o1-preview}      & 0.7583 $\uparrow$7.21\% & 646.94 \\
    \: + \textit{Distilled from DeepSeek-R1}     & 0.7859 $\uparrow$11.1\% & 1667.56 \\
    \: + \textit{Distilled from (o1 + R1)}       & 0.7780 $\uparrow$10.0\% & 1615.54 \\
    \addlinespace[2pt]
    Llama-Primus-Merged                             & 0.7603 $\uparrow$7.49\% & 241.92 \\
    \: + \textit{Distilled from o1-preview}      & 0.7780 $\uparrow$10.0\% & 726.96 \\
    \: + \textit{Distilled from DeepSeek-R1}     & 0.8075 $\uparrow$14.2\% & 1483.94 \\
    \: + \textit{Distilled from (o1 + R1)}       & 0.8193 $\uparrow$\textbf{15.8\%} & 1467.40 \\
    \midrule
    o1-preview                                      & 0.8035 & 1054.91 \\
    DeepSeek-R1                                     & 0.8212 & 1229.32 \\
    DeepSeek-R1-Distill-Llama-8B & 0.7399 & 1542.10 \\

    \bottomrule
\end{tabular}
\caption{Effect of \textsc{Primus-Reasoning} fine-tuning (on o1-preview, DeepSeek-R1, and their combination), evaluated on CISSP. $\uparrow$ indicates the percentage improvement over Llama without CoT and in the 5-shot setting. The best improvement is highlighted in \textbf{bold}.}
\label{table:cissp-token-comparison}
\end{table}


We further distill Llama-Primus-Merged using \textsc{Primus-Reasoning} (Sec.\ref{sec:primus-reasoning}), a high-quality dataset of cybersecurity task reasoning steps obtained from o1-preview and DeepSeek-R1, to equip it with reasoning and self-reflection capabilities. This approach has been successfully demonstrated in previous work such as S1~\cite{muennighoff2025s1} and Sky-T1~\cite{sky_t1_2025}. Since \textsc{Primus-Reasoning} is constructed from CTI-Bench tasks, we exclude them from the evaluation and choose CISSP as a representative metric, as it also emphasizes reasoning rather than just factual recall. The results are presented in Tab.\ref{table:cissp-token-comparison}.

As shown in the table, both Llama-3.1-8B-Instruct and Llama-Primus-Merged improve with CoT over direct answer generation. Notably, Llama-Primus-Merged achieves the largest gain, even outperforming DeepSeek-R1-Distill-Llama-8B\footnote{\scriptsize \href{https://huggingface.co/deepseek-ai/DeepSeek-R1-Distill-Llama-8B}{https://huggingface.co/deepseek-ai/DeepSeek-R1-Distill-Llama-8B}} (0.7603 vs. 0.7399) with the fewest tokens,\linebreak suggesting stronger cybersecurity knowledge benefits reasoning. After fine-tuning on \textsc{Primus-Reasoning} (rows starting with +), token usage increases while accuracy further improves; distillation on the combined o1-preview and DeepSeek-R1 data achieves the largest improvement \textbf{\emph{(15.8\%)}}. Interestingly, comparing DeepSeek-R1-Distill-Llama-8B (0.7399) and Llama-3.1-8B-Instruct after distillation (0.7583 / 0.7859 / 0.7780) may suggest that domain-specific reasoning distillation yields better in-domain performance than general-domain distillation.

\section{Domain Calibration Analysis}

\begin{table}[t]
\footnotesize
\centering
\setlength{\tabcolsep}{0.2pt}
\begin{tabular}{
    >{\raggedright\arraybackslash}p{1.6cm}
    >{\centering\arraybackslash}p{1.7cm}
    >{\centering\arraybackslash}p{2.15cm}
    >{\centering\arraybackslash}p{2.15cm}
}
\toprule
\multirow{3}{*}{\textbf{Benchmark}}
  & \multicolumn{3}{c}{\textbf{ECE (\%)}} \\
\cmidrule(lr){2-4}
  & \makecell[c]{Llama-3.1-\\8B-Instruct}
  & \makecell[c]{Llama-Primus-\\Base}
  & \makecell[c]{Llama-Primus-\\Merged} \\
\midrule
CISSP         &  7.22 & 4.59 & 4.55 \\
CTI-MCQ       & 11.01 & 2.03 & 5.52 \\
CyberMetric   &  4.11 & 3.41 & 2.57 \\
\midrule
\textbf{Average}
              &  7.45
              & 3.34\(\downarrow\)55.17\%
              & 4.21\(\downarrow\)43.49\% \\
\bottomrule
\end{tabular}
\caption{Expected Calibration Error (ECE) across cybersecurity benchmarks (with 10 bins).}
\label{table:ece_primus}
\end{table}

\begin{table}[t]
\footnotesize
\centering
\setlength{\tabcolsep}{0.55pt}
\begin{tabular}{
    >{\raggedright\arraybackslash}p{1.75cm}
    >{\centering\arraybackslash}p{1.6cm}
    >{\centering\arraybackslash}p{2.1cm}
    >{\centering\arraybackslash}p{2.1cm}
}
\toprule
\textbf{Metric}
  & \makecell[c]{\textbf{Llama-3.1-}\\\textbf{8B-Instruct}}
  & \makecell[c]{\textbf{Llama-Primus-}\\\textbf{Base}}
  & \makecell[c]{\textbf{Llama-Primus-}\\\textbf{Merged}} \\
\midrule
Accuracy (\%) & 67.56 & 66.29 & 66.59 \\
ECE (\%)      &  5.99 &  6.07 &  5.56 \\
\bottomrule
\end{tabular}
\caption{Accuracy and ECE across models on MMLU.}
\label{table:mmlu_performance}
\vspace{-1em}
\end{table}

In cybersecurity applications, a model's confidence score is often a critical indicator for deciding whether to escalate issues for human intervention, such as sending alerts to security analysts. For this to work, the confidence score must accurately reflect the true accuracy. After multi-stage training in the cybersecurity domain, we found that our model had a significantly lower Expected Calibration Error (ECE) \cite{guo2017calibration} on cybersecurity-related questions. This suggests our model's confidence is more aligned with its actual accuracy. The ECE measures the average discrepancy between a model's confidence and its empirical accuracy.

Specifically, we re-evaluated the cybersecurity multiple-choice tasks (CISSP, CTI-MCQ, and CyberMetric). We took the token probability of the output answer (A/B/C/D) as the confidence score and calculated the ECE, as shown in Tab.\ref{table:ece_primus}. The ECE of our model on cybersecurity questions was reduced by \textbf{half}, indicating that the model is better calibrated and thus more reliable in practical applications, especially those involving confidence thresholds. Additionally, evaluation on general-domain questions (e.g., MMLU) \cite{hendryckstest2021} showed no significant change (see Tab.\ref{table:mmlu_performance}).

Recent work has sought to improve LLM calibration by reducing ECE through specialized training methods \cite{xu-etal-2024-sayself}. However, leveraging domain-specific data for this purpose remains unexplored. We posit that our approach could provide valuable insights into confidence calibration.

\section{Conclusion}
In this work, we explore adapting other successful domain-specific LLM approaches to cybersecurity and contribute a series of datasets covering different stages of LLM training, including pre-training, instruction fine-tuning, and reasoning distillation, each of which has been validated to improve cybersecurity performance. To our knowledge, this is the \textbf{\emph{first}} study to systematically strengthen the cybersecurity skills of an LLM across multiple stages of training, and we will release all datasets and models to encourage further community research.

\section*{Limitations}
Although this work covers the various stages of LLM training, it has the following limitations:

\vspace{0.2\baselineskip}
\noindent\textbullet~Due to limited computational resources, our experiments primarily focus on 8B-scale models, leaving the effectiveness of scaling to larger models (e.g., 405B or 671B) unknown.

\vspace{0.2\baselineskip}
\noindent\textbullet~Our exploration of RL remains limited. Recent work by DeepSeek-R1 has demonstrated that GRPO \cite{zhang2024deepseekmath} combined with only rule-based rewards (e.g., correctness and format compliance) can achieve performance comparable to o1. We believe this is also a promising direction for cybersecurity applications and leave it as future work.

\section*{Ethics Statement}
We used Garak \cite{garak}, a toolkit that probes for hallucination, data leakage, prompt injection, misinformation, toxicity generation, jailbreaks, and many other vulnerabilities, to evaluate Llama-Primus-Merged. The results showed no significant differences compared to Llama (Appx.\ref{sec:appendix-safety}). However, we still emphasize that the user is solely responsible for the content generated with the Primus model, as it lacks mechanisms to handle the disclosure of harmful, biased, or toxic content. Therefore, we strongly recommend that Primus be used for research purposes only. If used in production for natural language generation, users should independently assess the risks and implement appropriate safeguards.

\bibliography{custom}

\begin{thebibliography}{62}
\providecommand{\natexlab}[1]{#1}

\bibitem[{Aghaei et~al.(2022)Aghaei, Niu, Shadid, and Al-Shaer}]{aghaei2022securebert}
Ehsan Aghaei, Xi~Niu, Waseem Shadid, and Ehab Al-Shaer. 2022.
\newblock Securebert: A domain-specific language model for cybersecurity.
\newblock In \emph{International Conference on Security and Privacy in Communication Systems}, pages 39--56. Springer.

\bibitem[{Alam et~al.(2024)Alam, Bhusal, Nguyen, and Rastogi}]{Alam2024CTIBench}
Md~Tanvirul Alam, Dipkamal Bhusal, Le~Nguyen, and Nidhi Rastogi. 2024.
\newblock \href {https://proceedings.neurips.cc/paper_files/paper/2024/hash/5acd3c628aa1819fbf07c39ef73e7285-Abstract-Datasets_and_Benchmarks_Track.html} {{CTIBench}: A benchmark for evaluating {LLMs} in cyber threat intelligence}.
\newblock In \emph{Advances in Neural Information Processing Systems 37 (NeurIPS 2024), Datasets and Benchmarks Track}.

\bibitem[{Anthropic(2024)}]{anthropic2024claude35sonnet}
Anthropic. 2024.
\newblock \href {https://www.anthropic.com/news/claude-3-5-sonnet} {Introducing claude 3.5 sonnet}.
\newblock Accessed: 2025-02-13.

\bibitem[{Brown et~al.(2020)Brown, Mann, Ryder, Subbiah, Kaplan, Dhariwal, Neelakantan, Shyam, Sastry, Askell, Agarwal, Herbert-Voss, Krueger, Henighan, Child, Ramesh, Ziegler, Wu, Winter, Hesse, Chen, Sigler, Litwin, Gray, Chess, Clark, Berner, McCandlish, Radford, Sutskever, and Amodei}]{10.5555/3495724.3495883}
Tom~B. Brown, Benjamin Mann, Nick Ryder, Melanie Subbiah, Jared Kaplan, Prafulla Dhariwal, Arvind Neelakantan, Pranav Shyam, Girish Sastry, Amanda Askell, Sandhini Agarwal, Ariel Herbert-Voss, Gretchen Krueger, Tom Henighan, Rewon Child, Aditya Ramesh, Daniel~M. Ziegler, Jeffrey Wu, Clemens Winter, Christopher Hesse, Mark Chen, Eric Sigler, Mateusz Litwin, Scott Gray, Benjamin Chess, Jack Clark, Christopher Berner, Sam McCandlish, Alec Radford, Ilya Sutskever, and Dario Amodei. 2020.
\newblock Language models are few-shot learners.
\newblock In \emph{Proceedings of the 34th International Conference on Neural Information Processing Systems}, NIPS '20, Red Hook, NY, USA. Curran Associates Inc.

\bibitem[{Colombo et~al.(2024)Colombo, Pires, Boudiaf, Culver, Melo, Corro, Martins, Esposito, Raposo, Morgado et~al.}]{colombo2024saullm}
Pierre Colombo, Telmo~Pessoa Pires, Malik Boudiaf, Dominic Culver, Rui Melo, Caio Corro, Andre~FT Martins, Fabrizio Esposito, Vera~L{\'u}cia Raposo, Sofia Morgado, et~al. 2024.
\newblock Saullm-7b: A pioneering large language model for law.
\newblock \emph{arXiv preprint arXiv:2403.03883}.

\bibitem[{{Common Crawl}(2008)}]{commoncrawl}
{Common Crawl}. 2008.
\newblock Common crawl.
\newblock \url{https://commoncrawl.org/}.
\newblock Accessed: 2025-02-13.

\bibitem[{Derczynski et~al.(2024)Derczynski, Galinkin, Martin, Majumdar, and Inie}]{garak}
Leon Derczynski, Erick Galinkin, Jeffrey Martin, Subho Majumdar, and Nanna Inie. 2024.
\newblock {garak: A Framework for Security Probing Large Language Models}.
\newblock \url{https://garak.ai}.
\newblock Accessed: 2025-02-16.

\bibitem[{Devlin et~al.(2019)Devlin, Chang, Lee, and Toutanova}]{kenton2019bert}
Jacob Devlin, Ming-Wei Chang, Kenton Lee, and Kristina Toutanova. 2019.
\newblock Bert: Pre-training of deep bidirectional transformers for language understanding.
\newblock In \emph{Proceedings of NAACL-HLT}, volume~1. Association for Computational Linguistics.

\bibitem[{Dubey et~al.(2024)Dubey, Jauhri, Pandey, Kadian, Al-Dahle, Letman, Mathur, Schelten, Yang, Fan et~al.}]{dubey2024llama}
Abhimanyu Dubey, Abhinav Jauhri, Abhinav Pandey, Abhishek Kadian, Ahmad Al-Dahle, Aiesha Letman, Akhil Mathur, Alan Schelten, Amy Yang, Angela Fan, et~al. 2024.
\newblock The llama 3 herd of models.
\newblock \emph{arXiv preprint arXiv:2407.21783}.

\bibitem[{Ferrag et~al.(2024)Ferrag, Alwahedi, Battah, Cherif, Mechri, and Tihanyi}]{ferrag2024generative}
Mohamed~Amine Ferrag, Fatima Alwahedi, Ammar Battah, Bilel Cherif, Abdechakour Mechri, and Norbert Tihanyi. 2024.
\newblock Generative ai and large language models for cyber security: All insights you need.
\newblock \emph{Available at SSRN 4853709}.

\bibitem[{Gao et~al.(2024)Gao, Tow, Abbasi, Biderman, Black, DiPofi, Foster, Golding, Hsu, Le~Noac'h, Li, McDonell, Muennighoff, Ociepa, Phang, Reynolds, Schoelkopf, Skowron, Sutawika, Tang, Thite, Wang, Wang, and Zou}]{eval-harness}
Leo Gao, Jonathan Tow, Baber Abbasi, Stella Biderman, Sid Black, Anthony DiPofi, Charles Foster, Laurence Golding, Jeffrey Hsu, Alain Le~Noac'h, Haonan Li, Kyle McDonell, Niklas Muennighoff, Chris Ociepa, Jason Phang, Laria Reynolds, Hailey Schoelkopf, Aviya Skowron, Lintang Sutawika, Eric Tang, Anish Thite, Ben Wang, Kevin Wang, and Andy Zou. 2024.
\newblock \href {https://doi.org/10.5281/zenodo.12608602} {A framework for few-shot language model evaluation}.

\bibitem[{Ge et~al.(2023)Ge, Hua, Mei, Tan, Xu, Li, Zhang et~al.}]{ge2023openagi}
Yingqiang Ge, Wenyue Hua, Kai Mei, Juntao Tan, Shuyuan Xu, Zelong Li, Yongfeng Zhang, et~al. 2023.
\newblock Openagi: When llm meets domain experts.
\newblock \emph{Advances in Neural Information Processing Systems}, 36:5539--5568.

\bibitem[{Gekhman et~al.(2024)Gekhman, Yona, Aharoni, Eyal, Feder, Reichart, and Herzig}]{gekhman-etal-2024-fine}
Zorik Gekhman, Gal Yona, Roee Aharoni, Matan Eyal, Amir Feder, Roi Reichart, and Jonathan Herzig. 2024.
\newblock \href {https://doi.org/10.18653/v1/2024.emnlp-main.444} {Does fine-tuning {LLM}s on new knowledge encourage hallucinations?}
\newblock In \emph{Proceedings of the 2024 Conference on Empirical Methods in Natural Language Processing}, pages 7765--7784, Miami, Florida, USA. Association for Computational Linguistics.

\bibitem[{Ghelani(2022)}]{ghelani2022cyber}
Diptiben Ghelani. 2022.
\newblock Cyber security, cyber threats, implications and future perspectives: A review.
\newblock \emph{Authorea Preprints}.

\bibitem[{Guo et~al.(2017)Guo, Pleiss, Sun, and Weinberger}]{guo2017calibration}
Chuan Guo, Geoff Pleiss, Yu~Sun, and Kilian~Q Weinberger. 2017.
\newblock On calibration of modern neural networks.
\newblock In \emph{International conference on machine learning}, pages 1321--1330. PMLR.

\bibitem[{Guo et~al.(2025)Guo, Yang, Zhang, Song, Zhang, Xu, Zhu, Ma, Wang, Bi et~al.}]{guo2025deepseek}
Daya Guo, Dejian Yang, Haowei Zhang, Junxiao Song, Ruoyu Zhang, Runxin Xu, Qihao Zhu, Shirong Ma, Peiyi Wang, Xiao Bi, et~al. 2025.
\newblock Deepseek-r1: Incentivizing reasoning capability in llms via reinforcement learning.
\newblock \emph{arXiv preprint arXiv:2501.12948}.

\bibitem[{Gururangan et~al.(2020)Gururangan, Marasovi{\'c}, Swayamdipta, Lo, Beltagy, Downey, and Smith}]{gururangan-etal-2020-dont}
Suchin Gururangan, Ana Marasovi{\'c}, Swabha Swayamdipta, Kyle Lo, Iz~Beltagy, Doug Downey, and Noah~A. Smith. 2020.
\newblock \href {https://doi.org/10.18653/v1/2020.acl-main.740} {Don`t stop pretraining: Adapt language models to domains and tasks}.
\newblock In \emph{Proceedings of the 58th Annual Meeting of the Association for Computational Linguistics}, pages 8342--8360, Online. Association for Computational Linguistics.

\bibitem[{Heafield(2011)}]{heafield-2011-kenlm}
Kenneth Heafield. 2011.
\newblock \href {https://aclanthology.org/W11-2123/} {{K}en{LM}: Faster and smaller language model queries}.
\newblock In \emph{Proceedings of the Sixth Workshop on Statistical Machine Translation}, pages 187--197, Edinburgh, Scotland. Association for Computational Linguistics.

\bibitem[{Hendrycks et~al.(2021)Hendrycks, Burns, Basart, Zou, Mazeika, Song, and Steinhardt}]{hendryckstest2021}
Dan Hendrycks, Collin Burns, Steven Basart, Andy Zou, Mantas Mazeika, Dawn Song, and Jacob Steinhardt. 2021.
\newblock Measuring massive multitask language understanding.
\newblock \emph{Proceedings of the International Conference on Learning Representations (ICLR)}.

\bibitem[{Huang et~al.(2024)Huang, Zou, Li, Liu, Zheng, Chern, Xia, Qin, Yuan, and Liu}]{huang2024o1}
Zhen Huang, Haoyang Zou, Xuefeng Li, Yixiu Liu, Yuxiang Zheng, Ethan Chern, Shijie Xia, Yiwei Qin, Weizhe Yuan, and Pengfei Liu. 2024.
\newblock O1 replication journey--part 2: Surpassing o1-preview through simple distillation, big progress or bitter lesson?
\newblock \emph{arXiv preprint arXiv:2411.16489}.

\bibitem[{Jackaduma(2021)}]{secbbert_github}
Jackaduma. 2021.
\newblock Secbert: A pretrained language model for cyber security text.
\newblock \url{https://github.com/jackaduma/SecBERT/}.
\newblock Accessed: 2025-02-03.

\bibitem[{Jiang et~al.(2023)Jiang, Wang, Liu, Xu, Tan, and Zhang}]{jiang2023nova}
Nan Jiang, Chengxiao Wang, Kevin Liu, Xiangzhe Xu, Lin Tan, and Xiangyu Zhang. 2023.
\newblock Nova: Generative language models for binaries.
\newblock \emph{arXiv preprint arXiv:2311.13721}.

\bibitem[{Jiao et~al.(2020)Jiao, Yin, Shang, Jiang, Chen, Li, Wang, and Liu}]{jiao-etal-2020-tinybert}
Xiaoqi Jiao, Yichun Yin, Lifeng Shang, Xin Jiang, Xiao Chen, Linlin Li, Fang Wang, and Qun Liu. 2020.
\newblock \href {https://doi.org/10.18653/v1/2020.findings-emnlp.372} {{T}iny{BERT}: Distilling {BERT} for natural language understanding}.
\newblock In \emph{Findings of the Association for Computational Linguistics: EMNLP 2020}, pages 4163--4174, Online. Association for Computational Linguistics.

\bibitem[{Labrak et~al.(2024)Labrak, Bazoge, Morin, Gourraud, Rouvier, and Dufour}]{labrak-etal-2024-biomistral}
Yanis Labrak, Adrien Bazoge, Emmanuel Morin, Pierre-Antoine Gourraud, Mickael Rouvier, and Richard Dufour. 2024.
\newblock \href {https://doi.org/10.18653/v1/2024.findings-acl.348} {{B}io{M}istral: A collection of open-source pretrained large language models for medical domains}.
\newblock In \emph{Findings of the Association for Computational Linguistics: ACL 2024}, pages 5848--5864, Bangkok, Thailand. Association for Computational Linguistics.

\bibitem[{Lai et~al.(2024)Lai, Gan, Wu, Qi, and Philip}]{lai2024large}
Jinqi Lai, Wensheng Gan, Jiayang Wu, Zhenlian Qi, and S~Yu Philip. 2024.
\newblock Large language models in law: A survey.
\newblock \emph{AI Open}.

\bibitem[{Lee et~al.(2022)Lee, Ippolito, Nystrom, Zhang, Eck, Callison-Burch, and Carlini}]{lee-etal-2022-deduplicating}
Katherine Lee, Daphne Ippolito, Andrew Nystrom, Chiyuan Zhang, Douglas Eck, Chris Callison-Burch, and Nicholas Carlini. 2022.
\newblock \href {https://doi.org/10.18653/v1/2022.acl-long.577} {Deduplicating training data makes language models better}.
\newblock In \emph{Proceedings of the 60th Annual Meeting of the Association for Computational Linguistics (Volume 1: Long Papers)}, pages 8424--8445, Dublin, Ireland. Association for Computational Linguistics.

\bibitem[{Li et~al.(2023{\natexlab{a}})Li, Li, Guannan, Yang, and Yu}]{li2023seceval}
Guancheng Li, Yifeng Li, Wang Guannan, Haoyu Yang, and Yang Yu. 2023{\natexlab{a}}.
\newblock Seceval: A comprehensive benchmark for evaluating cybersecurity knowledge of foundation models.
\newblock https://github.com/XuanwuAI/SecEval.

\bibitem[{Li et~al.(2023{\natexlab{b}})Li, Bubeck, Eldan, Del~Giorno, Gunasekar, and Lee}]{li2023textbooks}
Yuanzhi Li, S{\'e}bastien Bubeck, Ronen Eldan, Allie Del~Giorno, Suriya Gunasekar, and Yin~Tat Lee. 2023{\natexlab{b}}.
\newblock Textbooks are all you need ii: phi-1.5 technical report.
\newblock \emph{arXiv preprint arXiv:2309.05463}.

\bibitem[{Li and Liu(2021)}]{li2021comprehensive}
Yuchong Li and Qinghui Liu. 2021.
\newblock A comprehensive review study of cyber-attacks and cyber security; emerging trends and recent developments.
\newblock \emph{Energy Reports}, 7:8176--8186.

\bibitem[{Liu et~al.(2024)Liu, Feng, Xue, Wang, Wu, Lu, Zhao, Deng, Zhang, Ruan et~al.}]{liu2024deepseek}
Aixin Liu, Bei Feng, Bing Xue, Bingxuan Wang, Bochao Wu, Chengda Lu, Chenggang Zhao, Chengqi Deng, Chenyu Zhang, Chong Ruan, et~al. 2024.
\newblock Deepseek-v3 technical report.
\newblock \emph{arXiv preprint arXiv:2412.19437}.

\bibitem[{Minaee et~al.(2024)Minaee, Mikolov, Nikzad, Chenaghlu, Socher, Amatriain, and Gao}]{minaee2024large}
Shervin Minaee, Tomas Mikolov, Narjes Nikzad, Meysam Chenaghlu, Richard Socher, Xavier Amatriain, and Jianfeng Gao. 2024.
\newblock Large language models: A survey.
\newblock \emph{arXiv preprint arXiv:2402.06196}.

\bibitem[{Mosaic(2024)}]{Mosaic2024DBRX}
Mosaic. 2024.
\newblock \href {https://www.databricks.com/blog/introducing-dbrx-new-state-art-open-llm} {Introducing dbrx: A new state-of-the-art open llm}.
\newblock Accessed: 2025-02-13.

\bibitem[{Muennighoff et~al.(2025)Muennighoff, Yang, Shi, Li, Fei-Fei, Hajishirzi, Zettlemoyer, Liang, Cand{\`e}s, and Hashimoto}]{muennighoff2025s1}
Niklas Muennighoff, Zitong Yang, Weijia Shi, Xiang~Lisa Li, Li~Fei-Fei, Hannaneh Hajishirzi, Luke Zettlemoyer, Percy Liang, Emmanuel Cand{\`e}s, and Tatsunori Hashimoto. 2025.
\newblock s1: Simple test-time scaling.
\newblock \emph{arXiv preprint arXiv:2501.19393}.

\bibitem[{NVIDIA(2025)}]{NVIDIA_NeMo}
NVIDIA. 2025.
\newblock \href {https://github.com/NVIDIA/NeMo} {Nemo: A scalable generative ai framework}.
\newblock Accessed: 2025-02-13.

\bibitem[{Ouyang et~al.(2022)Ouyang, Wu, Jiang, Almeida, Wainwright, Mishkin, Zhang, Agarwal, Slama, Ray et~al.}]{ouyang2022training}
Long Ouyang, Jeffrey Wu, Xu~Jiang, Diogo Almeida, Carroll Wainwright, Pamela Mishkin, Chong Zhang, Sandhini Agarwal, Katarina Slama, Alex Ray, et~al. 2022.
\newblock Training language models to follow instructions with human feedback.
\newblock \emph{Advances in neural information processing systems}, 35:27730--27744.

\bibitem[{Park and You(2023)}]{park2023pretrained}
Youngja Park and Weiqiu You. 2023.
\newblock A pretrained language model for cyber threat intelligence.
\newblock In \emph{Proceedings of the 2023 Conference on Empirical Methods in Natural Language Processing: Industry Track}, pages 113--122.

\bibitem[{Penedo et~al.(2024)Penedo, Kydl{\'\i}{\v{c}}ek, allal, Lozhkov, Mitchell, Raffel, Werra, and Wolf}]{penedo2024the}
Guilherme Penedo, Hynek Kydl{\'\i}{\v{c}}ek, Loubna~Ben allal, Anton Lozhkov, Margaret Mitchell, Colin Raffel, Leandro~Von Werra, and Thomas Wolf. 2024.
\newblock \href {https://openreview.net/forum?id=n6SCkn2QaG} {The fineweb datasets: Decanting the web for the finest text data at scale}.
\newblock In \emph{The Thirty-eight Conference on Neural Information Processing Systems Datasets and Benchmarks Track}.

\bibitem[{Radford(2018)}]{radford2018improving}
Alec Radford. 2018.
\newblock Improving language understanding by generative pre-training.
\newblock \emph{OpenAI Blog}.

\bibitem[{Raffel et~al.(2019)Raffel, Shazeer, Roberts, Lee, Narang, Matena, Zhou, Li, and Liu}]{raffel2019exploringc4}
Colin Raffel, Noam Shazeer, Adam Roberts, Katherine Lee, Sharan Narang, Michael Matena, Yanqi Zhou, Wei Li, and Peter~J Liu. 2019.
\newblock Exploring the limits of transfer learning with a unified text-to-text transformer.
\newblock \emph{arXiv preprint arXiv:1910.10683}.

\bibitem[{Ranade et~al.(2021)Ranade, Piplai, Joshi, and Finin}]{ranade2021cybert}
Priyanka Ranade, Aritran Piplai, Anupam Joshi, and Tim Finin. 2021.
\newblock Cybert: Contextualized embeddings for the cybersecurity domain.
\newblock In \emph{2021 IEEE International Conference on Big Data (Big Data)}, pages 3334--3342. IEEE.

\bibitem[{Su et~al.(2024)Su, Kong, Lin, Jennings, Norick, Kliegl, Patwary, Shoeybi, and Catanzaro}]{su2024nemotroncctransformingcommoncrawl}
Dan Su, Kezhi Kong, Ying Lin, Joseph Jennings, Brandon Norick, Markus Kliegl, Mostofa Patwary, Mohammad Shoeybi, and Bryan Catanzaro. 2024.
\newblock \href {https://arxiv.org/abs/2412.02595} {Nemotron-cc: Transforming common crawl into a refined long-horizon pretraining dataset}.
\newblock \emph{Preprint}, arXiv:2412.02595.

\bibitem[{Sun et~al.(2019)Sun, Ho, and yi~Lee}]{Sun2019LAMOLLM}
Fan-Keng Sun, Cheng-Hao Ho, and Hung yi~Lee. 2019.
\newblock \href {https://api.semanticscholar.org/CorpusID:209475822} {Lamol: Language modeling for lifelong language learning}.
\newblock In \emph{International Conference on Learning Representations}.

\bibitem[{Sun et~al.(2023)Sun, Allix, Kim, Zhou, Kim, Lo, Bissyand{\'e}, and Klein}]{sun2023dexbert}
Tiezhu Sun, Kevin Allix, Kisub Kim, Xin Zhou, Dongsun Kim, David Lo, Tegawend{\'e}~F Bissyand{\'e}, and Jacques Klein. 2023.
\newblock Dexbert: Effective, task-agnostic and fine-grained representation learning of android bytecode.
\newblock \emph{IEEE Transactions on Software Engineering}.

\bibitem[{Team(2025)}]{sky_t1_2025}
NovaSky Team. 2025.
\newblock Sky-t1: Train your own o1 preview model within \$450.
\newblock https://novasky-ai.github.io/posts/sky-t1.
\newblock Accessed: 2025-01-09.

\bibitem[{Tihanyi et~al.(2024)Tihanyi, Ferrag, Jain, Bisztray, and Debbah}]{cybermetric}
Norbert Tihanyi, Mohamed~Amine Ferrag, Ridhi Jain, Tamas Bisztray, and Merouane Debbah. 2024.
\newblock \href {https://doi.org/10.1109/CSR61664.2024.10679494} {Cybermetric: A benchmark dataset based on retrieval-augmented generation for evaluating llms in cybersecurity knowledge}.
\newblock In \emph{2024 IEEE International Conference on Cyber Security and Resilience (CSR)}, pages 296--302.

\bibitem[{Touvron et~al.(2023)Touvron, Lavril, Izacard, Martinet, Lachaux, Lacroix, Rozi{\`e}re, Goyal, Hambro, Azhar et~al.}]{touvron2023llama}
Hugo Touvron, Thibaut Lavril, Gautier Izacard, Xavier Martinet, Marie-Anne Lachaux, Timoth{\'e}e Lacroix, Baptiste Rozi{\`e}re, Naman Goyal, Eric Hambro, Faisal Azhar, et~al. 2023.
\newblock Llama: Open and efficient foundation language models.
\newblock \emph{arXiv preprint arXiv:2302.13971}.

\bibitem[{Wang et~al.(2024{\natexlab{a}})Wang, Gao, Zhang, Sha, Sun, Zhou, Zhu, Sun, Qiu, and Xiao}]{wang2024clap}
Hao Wang, Zeyu Gao, Chao Zhang, Zihan Sha, Mingyang Sun, Yuchen Zhou, Wenyu Zhu, Wenju Sun, Han Qiu, and Xi~Xiao. 2024{\natexlab{a}}.
\newblock Clap: Learning transferable binary code representations with natural language supervision.
\newblock In \emph{Proceedings of the 33rd ACM SIGSOFT International Symposium on Software Testing and Analysis}, pages 503--515.

\bibitem[{Wang et~al.(2024{\natexlab{b}})Wang, Zhang, Zhang, Hu, Li, Zhang, Li, Wu, Wang, and Hovy}]{wang2024reinforcement}
Shuhe Wang, Shengyu Zhang, Jie Zhang, Runyi Hu, Xiaoya Li, Tianwei Zhang, Jiwei Li, Fei Wu, Guoyin Wang, and Eduard Hovy. 2024{\natexlab{b}}.
\newblock Reinforcement learning enhanced llms: A survey.
\newblock \emph{arXiv preprint arXiv:2412.10400}.

\bibitem[{Wei et~al.(2022)Wei, Tay, Bommasani, Raffel, Zoph, Borgeaud, Yogatama, Bosma, Zhou, Metzler et~al.}]{wei2022emergent}
Jason Wei, Yi~Tay, Rishi Bommasani, Colin Raffel, Barret Zoph, Sebastian Borgeaud, Dani Yogatama, Maarten Bosma, Denny Zhou, Donald Metzler, et~al. 2022.
\newblock Emergent abilities of large language models.
\newblock \emph{arXiv preprint arXiv:2206.07682}.

\bibitem[{Wenzek et~al.(2020)Wenzek, Lachaux, Conneau, Chaudhary, Guzm{\'a}n, Joulin, and Grave}]{wenzek-etal-2020-ccnet}
Guillaume Wenzek, Marie-Anne Lachaux, Alexis Conneau, Vishrav Chaudhary, Francisco Guzm{\'a}n, Armand Joulin, and Edouard Grave. 2020.
\newblock \href {https://aclanthology.org/2020.lrec-1.494/} {{CCN}et: Extracting high quality monolingual datasets from web crawl data}.
\newblock In \emph{Proceedings of the Twelfth Language Resources and Evaluation Conference}, pages 4003--4012, Marseille, France. European Language Resources Association.

\bibitem[{Xu et~al.(2024{\natexlab{a}})Xu, Wang, Li, Wang, Zhao, Chen, Yu, Liu, and Wang}]{xu2024large}
HanXiang Xu, ShenAo Wang, Ningke Li, Kailong Wang, Yanjie Zhao, Kai Chen, Ting Yu, Yang Liu, and HaoYu Wang. 2024{\natexlab{a}}.
\newblock Large language models for cyber security: A systematic literature review.
\newblock \emph{arXiv preprint arXiv:2405.04760}.

\bibitem[{Xu et~al.(2024{\natexlab{b}})Xu, Wu, Diao, Liu, Wang, Chen, and Gao}]{xu-etal-2024-sayself}
Tianyang Xu, Shujin Wu, Shizhe Diao, Xiaoze Liu, Xingyao Wang, Yangyi Chen, and Jing Gao. 2024{\natexlab{b}}.
\newblock \href {https://doi.org/10.18653/v1/2024.emnlp-main.343} {{S}ay{S}elf: Teaching {LLM}s to express confidence with self-reflective rationales}.
\newblock In \emph{Proceedings of the 2024 Conference on Empirical Methods in Natural Language Processing}, pages 5985--5998, Miami, Florida, USA. Association for Computational Linguistics.

\bibitem[{Yadav et~al.(2023)Yadav, Tam, Choshen, Raffel, and Bansal}]{yadav2023tiesmerging}
Prateek Yadav, Derek Tam, Leshem Choshen, Colin Raffel, and Mohit Bansal. 2023.
\newblock \href {https://proceedings.neurips.cc/paper_files/paper/2023/hash/1644c9af28ab7916874f6fd6228a9bcf-Abstract-Conference.html} {{TIES}-merging: Resolving interference when merging models}.
\newblock In \emph{Advances in Neural Information Processing Systems 36 (NeurIPS 2023)}.

\bibitem[{Yan et~al.(2024)Yan, Sha, Zhao, Li, Martinez-Maldonado, Chen, Li, Jin, and Ga{\v{s}}evi{\'c}}]{yan2024practical}
Lixiang Yan, Lele Sha, Linxuan Zhao, Yuheng Li, Roberto Martinez-Maldonado, Guanliang Chen, Xinyu Li, Yueqiao Jin, and Dragan Ga{\v{s}}evi{\'c}. 2024.
\newblock Practical and ethical challenges of large language models in education: A systematic scoping review.
\newblock \emph{British Journal of Educational Technology}, 55(1):90--112.

\bibitem[{Yu et~al.(2024)Yu, Yu, Yu, Huang, and Li}]{yu2024languageDARES}
Le~Yu, Bowen Yu, Haiyang Yu, Fei Huang, and Yongbin Li. 2024.
\newblock \href {https://arxiv.org/abs/2311.03099} {Language models are super mario: Absorbing abilities from homologous models as a free lunch}.
\newblock In \emph{Proceedings of the 41st International Conference on Machine Learning (ICML)}. PMLR.

\bibitem[{Zhang et~al.(2024{\natexlab{a}})Zhang, Bu, Wen, Chen, Li, and Zhu}]{zhang2024llms}
Jie Zhang, Haoyu Bu, Hui Wen, Yu~Chen, Lun Li, and Hongsong Zhu. 2024{\natexlab{a}}.
\newblock When llms meet cybersecurity: A systematic literature review.
\newblock \emph{arXiv preprint arXiv:2405.03644}.

\bibitem[{Zhang et~al.(2023{\natexlab{a}})Zhang, Wen, Deng, Xin, Li, Li, Zhu, and Sun}]{hackmentor2023}
Jie Zhang, Hui Wen, Liting Deng, Mingfeng Xin, Zhi Li, Lun Li, Hongsong Zhu, and Limin Sun. 2023{\natexlab{a}}.
\newblock \href {https://doi.org/10.1109/TrustCom60117.2023.00076} {Hackmentor: Fine-tuning large language models for cybersecurity}.
\newblock In \emph{2023 IEEE 22nd International Conference on Trust, Security and Privacy in Computing and Communications (TrustCom)}, pages 452--461.

\bibitem[{Zhang et~al.(2023{\natexlab{b}})Zhang, Dong, Li, Zhang, Sun, Wang, Li, Hu, Zhang, Wu et~al.}]{zhang2023instruction}
Shengyu Zhang, Linfeng Dong, Xiaoya Li, Sen Zhang, Xiaofei Sun, Shuhe Wang, Jiwei Li, Runyi Hu, Tianwei Zhang, Fei Wu, et~al. 2023{\natexlab{b}}.
\newblock Instruction tuning for large language models: A survey.
\newblock \emph{arXiv preprint arXiv:2308.10792}.

\bibitem[{Zhang et~al.(2024{\natexlab{b}})Zhang, Li, Wang, and Liu}]{zhang2024deepseekmath}
Wei Zhang, Ming Li, Hao Wang, and Yang Liu. 2024{\natexlab{b}}.
\newblock \href {https://arxiv.org/abs/2402.03300} {Deepseekmath: Scalable math pre-training and group relative policy optimization for mathematical reasoning}.
\newblock \emph{arXiv preprint arXiv:2402.03300}.

\bibitem[{Zheng et~al.(2023)Zheng, Chiang, Sheng, Zhuang, Wu, Zhuang, Lin, Li, Li, Xing, Zhang, Gonzalez, and Stoica}]{zheng2023judging}
Lianmin Zheng, Wei-Lin Chiang, Ying Sheng, Siyuan Zhuang, Zhanghao Wu, Yonghao Zhuang, Zi~Lin, Zhuohan Li, Dacheng Li, Eric~P. Xing, Hao Zhang, Joseph~E. Gonzalez, and Ion Stoica. 2023.
\newblock \href {https://proceedings.neurips.cc/paper_files/paper/2023/hash/91f18a1287b398d378ef22505bf41832-Abstract-Datasets_and_Benchmarks.html} {Judging llm-as-a-judge with mt-bench and chatbot arena}.
\newblock In \emph{Advances in Neural Information Processing Systems 36 (NeurIPS 2023), Datasets and Benchmarks Track}.

\bibitem[{Zheng et~al.(2024)Zheng, Zhang, Zhang, Ye, Luo, Feng, and Ma}]{zheng2024llamafactory}
Yaowei Zheng, Richong Zhang, Junhao Zhang, Yanhan Ye, Zheyan Luo, Zhangchi Feng, and Yongqiang Ma. 2024.
\newblock \href {http://arxiv.org/abs/2403.13372} {Llamafactory: Unified efficient fine-tuning of 100+ language models}.
\newblock In \emph{Proceedings of the 62nd Annual Meeting of the Association for Computational Linguistics (Volume 3: System Demonstrations)}, Bangkok, Thailand. Association for Computational Linguistics.

\bibitem[{Zhou et~al.(2023)Zhou, Liu, Gu, Zou, Huang, Wu, Li, Chen, Zhou, Liu et~al.}]{zhou2023survey}
Hongjian Zhou, Fenglin Liu, Boyang Gu, Xinyu Zou, Jinfa Huang, Jinge Wu, Yiru Li, Sam~S Chen, Peilin Zhou, Junling Liu, et~al. 2023.
\newblock A survey of large language models in medicine: Progress, application, and challenge.
\newblock \emph{arXiv preprint arXiv:2311.05112}.

\end{thebibliography}

\appendix
\clearpage
\section{FAQs}
\label{sec:appendix-faqs}
\noindent\textbullet~\textbf{Q1}: \textbf{\emph{What are the implementation details, such as the training hyperparameters and the prompts used for the LLM during dataset construction?}}

These details are provided in the appendix. The training hyperparameters are listed in Appx.\ref{sec:appendix-hyperparameters}, and the prompts used for dataset construction are included in Appx.\ref{sec:appendix-prompts}.

\vspace{0.4\baselineskip}
\noindent\textbullet~\textbf{Q2}: \textbf{\emph{The experiments primarily target 8B models. A natural follow-up is whether these datasets generalize to larger models, i.e., whether they can also improve the cybersecurity performance of larger models?}}

\begin{table*}[t]
\footnotesize
\centering
\setlength{\tabcolsep}{1.0pt} 
\begin{tabular}{>{\raggedright\arraybackslash}p{4.6cm} cccccccc}
    \toprule
    \textbf{Model} & \textbf{CISSP} & \textbf{CTI-MCQ} & \textbf{CTI-RCM} & \textbf{CTI-VSP} & \textbf{CTI-ATE} & \textbf{CyberMetric} & \textbf{SecEval} & \textbf{\textit{Agg.}} \\
    \midrule
    Llama-3.1-Nemotron-70B-Instruct        & 0.8527      & 0.6900      & 0.6590      & 1.1893      & 0.3905      & \textbf{0.9380} & 0.7177      & 3.06      \\
    Llama-Primus-Nemotron-70B-Base         & \textbf{0.8703} & \textbf{0.7148} & \textbf{0.7410} & \textbf{1.0281} & \textbf{0.4540} & 0.9280      & \textbf{0.7208} & \textbf{3.40$\uparrow$11.2\%} \\
    \bottomrule
\end{tabular}
\caption{Performance comparison of Llama-3.1-Nemotron-70B-Instruct and Llama-Primus-Nemotron-70B-Base on cybersecurity benchmarks. CTI-VSP is scored using Mean Absolute Deviation \textbf{\emph{(lower is better)}}, CTI-ATE uses F1 score, and the others use accuracy. The aggregate score \emph{(Agg.)} is the sum of all benchmarks, with CTI-VSP negated. The best results are highlighted in \textbf{bold}.}
\label{table:primus-nemotron-base-performance}
\end{table*}


Yes, we extended our experiments to a 70B model by further pre-training Llama-3.1-Nemotron-70B-Instruct to obtain Llama-Primus-Nemotron-70B-Base. In addition to the dataset used for the 8B model, we supplemented its pre-training corpus with 7.6B tokens of cybersecurity content filtered from Nemotron-CC \cite{su2024nemotroncctransformingcommoncrawl} (see Appx.\ref{sec:appendix-primus-nemotron-cc}). The results in Tab.\ref{table:primus-nemotron-base-performance} show an \textbf{11.2\%} gain in the aggregated cybersecurity benchmark score. We will also release this model under the MIT license. Due to its high computational cost, we did not conduct the dataset-combination ablation study on the 70B model that we performed on the 8B experiments.

\vspace{0.4\baselineskip}
\noindent\textbullet~\textbf{Q3}: \textbf{\emph{Since LLMs (e.g., Claude) were used during dataset construction, has their reliability been evaluated?}}

\begin{table}[t]
\footnotesize
\centering
\setlength{\tabcolsep}{0.1pt}
\begin{tabular}{
    >{\raggedright\arraybackslash}p{5.1cm}   
    >{\centering\arraybackslash}p{1.25cm}   
    >{\centering\arraybackslash}p{1.25cm}   
}
\toprule
\textbf{Task}
  & \makecell{\textbf{MAE}\\{(Claude)}}
  & \makecell{\textbf{MAE}\\{(GPT-4o)}} \\
\midrule
Alert Explanation                       & 0.8  & 1.0  \\
Retrieved Security Doc QA               & 0.7  & 1.1  \\
Suspicious Command Analysis             & 0.4  & 1.0  \\
Security Event Query Generation         & 1.0  & 0.8  \\
Terraform Security Misconfiguration Fix & 1.1  & 0.4  \\
\midrule
\textbf{Average}                        & 0.8  & 0.86 \\
\bottomrule
\end{tabular}
\caption{Mean absolute error (MAE) between human expert scores and LLM scores across different \textsc{Primus‐Instruct} tasks.}
\label{table:primus_instruct_alignment}
\end{table}

\begin{figure*}[t]
  \centering
  \includegraphics[width=0.85\textwidth]{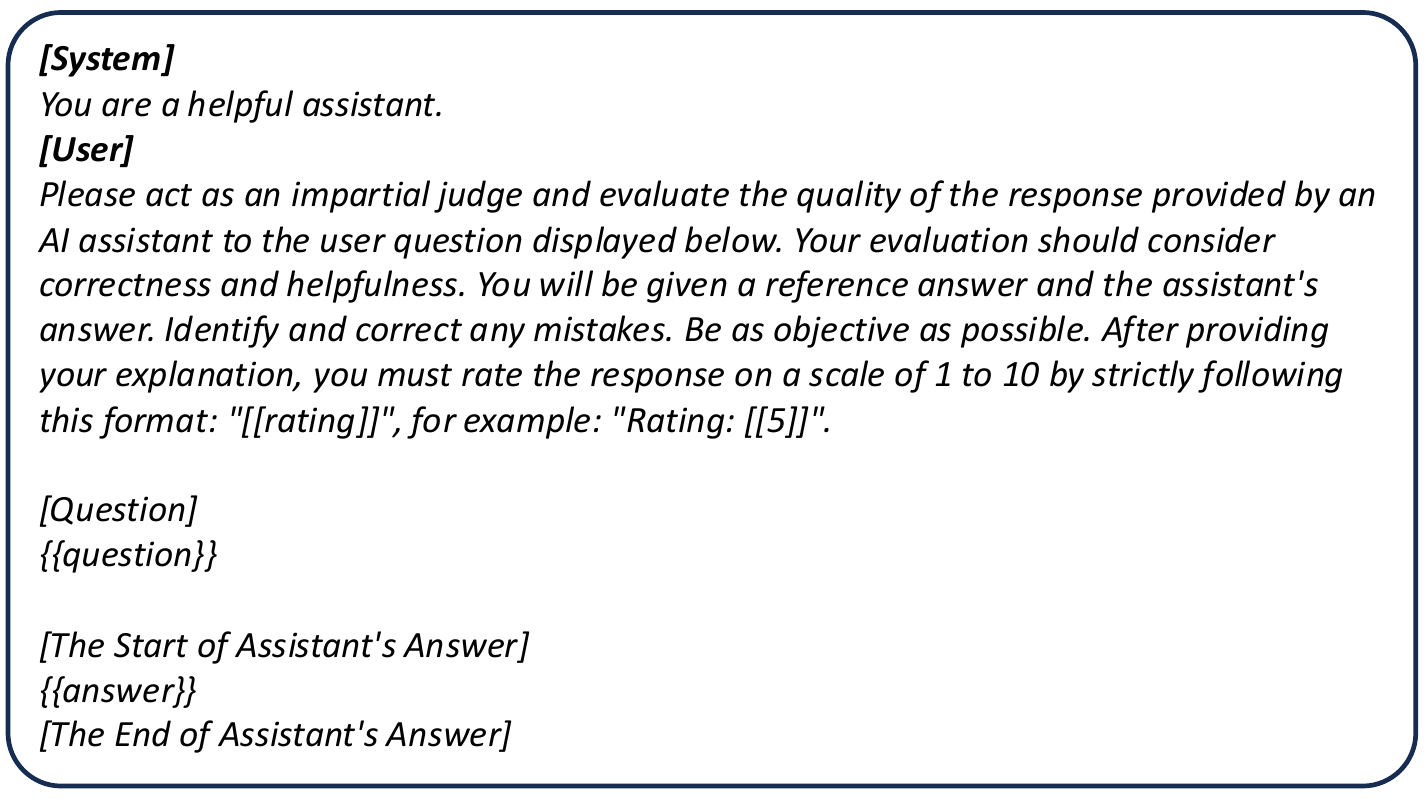}
  \caption{Judge prompt for evaluating response quality during \textsc{Primus-Instruct} generation.}
  \label{fig:prompt-primus-instruct-judge}
\end{figure*}

Yes, we conducted an experiment to measure the discrepancy between human experts and LLM judges under identical prompts. Specifically, in Sec.\ref{sec:primus-instruct} we used Claude 3.5 Sonnet to rate the helpfulness of responses in \textsc{Primus-Instruct}, discarding those that were not helpful enough\footnote{\scriptsize The judge prompt is provided in the Appx.\ref{sec:appendix-prompts} (Fig.\ref{fig:prompt-primus-instruct-judge})}. To validate Claude's reliability as a judge, we randomly selected ten examples per task for human experts to score, then computed the differences between human, GPT-4o, and Claude ratings.

The discrepancies are reported in Tab.\ref{table:primus_instruct_alignment}. Since \textsc{Primus-Instruct}'s responses were generated by GPT-4o, we found that it tended to favor its own answers, which is consistent with findings in LLM-as-a-Judge \cite{zheng2023judging}. This resulted in slightly larger discrepancies compared to Claude. Based on these results, we found that the gap between LLM-based and human scoring remained within an acceptable range.

\vspace{0.4\baselineskip}
\noindent\textbullet~\textbf{Q4}: \textbf{\emph{What is the training objective of}} \textsc{\textbf{Primus-Instruct}}\textbf{\emph{?}}

We would like to clarify that our primary goal with the SFT data was \textbf{\emph{not}} to further improve the model's cybersecurity capabilities. Instead, our goal was to help the model regain its instruction-following ability \textbf{\emph{without forgetting}} the cybersecurity knowledge acquired during pre-training. This can be viewed as a continual learning problem involving two tasks: "retaining cybersecurity knowledge" and "learning instruction following". According to LAMOL \cite{Sun2019LAMOLLM}, language models often suffer from catastrophic forgetting when trained sequentially on multiple tasks---learning a new task tends to overwrite knowledge from previous ones.

A common solution is to interleave data from previous tasks into the new task to mitigate forgetting. Inspired by this, we designed our cybersecurity SFT data to combine both instruction-following and domain-specific knowledge, hoping that the model would learn instruction-following while retaining its earlier cybersecurity understanding. As shown in Tab.\ref{table:primus-instruction-performance}, the results suggest that the model was able to recover instruction-following ability without significant loss in cybersecurity performance.

\vspace{0.4\baselineskip}
\noindent\textbullet~\textbf{Q5}: \textbf{\emph{Why does}} \textsc{\textbf{Primus-Instruct}} \textbf{\emph{appear to have a relatively small number of samples (\textasciitilde 1k)?}}

\begin{table}[t]
\footnotesize
\centering
\setlength{\tabcolsep}{0.1pt}
\begin{tabular}{
    >{\raggedright\arraybackslash}p{5.1cm}   
    >{\centering\arraybackslash}p{1.1cm}     
    >{\centering\arraybackslash}p{1.45cm}     
}
\toprule
\textbf{Task}
  & \textbf{Samples}
  & \textbf{Accepted} \\
\midrule
Alert Explanation                         &  400 & 100 \\
Retrieved Security Doc QA                 &  400 & 100 \\
Suspicious Command Analysis               &  400 & 100 \\
Security Event Query Generation           &  400 & 100 \\
Terraform Security Misconfiguration Fix   &  300 &  96 \\
\midrule
\textbf{Total}                            & 1,900 & 496 \\
\hspace{0.5mm}+ General Instruction Following (339) & 2,239 & 835 \\
\bottomrule
\end{tabular}
\caption{Initially designed (unfiltered) and accepted (filtered) sample counts per task, where accepted refers to the top 100 samples with a judge score $\ge 8$.}
\label{table:primus_instruct_sample_counts}
\end{table}

\begin{table*}[t]
\footnotesize
\centering
\setlength{\tabcolsep}{1pt}
\begin{tabular}{
    >{\raggedright\arraybackslash}p{3.0cm}
    >{\centering\arraybackslash}p{2.1cm}
    >{\centering\arraybackslash}p{3.5cm}
    >{\centering\arraybackslash}p{3.2cm}
    >{\centering\arraybackslash}p{2.0cm}
    >{\centering\arraybackslash}p{1.5cm}
}
\toprule
\textbf{Model}
  & \makecell[c]{\textbf{Base Model for}\\\textbf{Merge}}
  & \makecell[c]{\textbf{Merge Model 1}\\\textbf{(Task Vector 1)}}
  & \makecell[c]{\textbf{Merge Model 2}\\\textbf{(Task Vector 2)}}
  & \makecell[c]{\textbf{Cybersecurity}\\\textbf{\textit{Agg.}~Score}}
  & \textbf{MT-Bench} \\
\midrule
\makecell[l]{Llama-Primus-Merged\\(from unfiltered SFT)}
  & Llama-3.1-8b
  & \makecell[c]{Llama-Primus-Base\\-> SFT (2,239 samples)}
  & Llama-3.1-8b-Instruct
  & 2.44
  & 7.97 \\
\midrule
\makecell[l]{Llama-Primus-Merged\\(from filtered SFT)}
  & Llama-3.1-8b
  & \makecell[c]{Llama-Primus-Base\\-> SFT (835 samples)}
  & Llama-3.1-8b-Instruct
  & 2.63
  & 8.29 \\
\midrule
\makecell[l]{Llama-3.1-8b-Instruct}
  & -- & -- & -- & 2.29 & 8.35 \\
\bottomrule
\end{tabular}
\caption{Comparison of merged \textsc{Primus} models using different versions of the SFT dataset on cybersecurity and MT-Bench benchmarks. The first row refers to applying SFT on Llama-Primus-Base using the unfiltered 2,239 samples from Tab.\ref{table:primus_instruct_sample_counts} before merging with Llama-3.1-8B-Instruct, while the second row uses the filtered high-quality 835-sample version for SFT prior to merging.}

\label{tab:primus_merged_models}
\end{table*}

In fact, \textsc{Primus-Instruct} was selected from a larger pool of data. For each task, we initially generated 300–400 samples and rated their helpfulness (on a scale of 1 to 10) using the judge prompt in Fig.\ref{fig:prompt-primus-instruct-judge}. Only the top 100 samples with scores of at least 8 were retained (Tab.\ref{table:primus_instruct_sample_counts}).

Since we first performed SFT and then merged the resulting model with Llama-3.1-8B-Instruct to balance cybersecurity capabilities and instruction-following ability (Sec.\ref{sec:train-instruct-merge}), the \textbf{\emph{SFT and merging steps should be considered as a unified process}}. We therefore evaluated the combined effect of both. Specifically, we conducted SFT on Llama-Primus-Base separately using both the unfiltered version (2,239 samples) and the filtered high-quality version (835 samples) from Tab.\ref{table:primus_instruct_sample_counts}. Each resulting SFT
 model was then merged with Llama-3.1-8B-Instruct for comparison.

The merging process involves subtracting each model’s weights from the same base model (Llama-3.1-8B) to obtain two task vectors: one representing cybersecurity knowledge, and the other representing instruction-following ability. The results are shown in Tab.\ref{tab:primus_merged_models}. We found that applying SFT with a small amount (835) of high-quality data on Llama-Primus-Base before merging yields the best results in both the Cybersecurity Aggregate Score (2.63) and the MT-Bench score (8.29). This is why we chose the filtered high-quality version as \textsc{Primus-Instruct}.

\vspace{0.4\baselineskip}
\noindent\textbullet~\textbf{Q6}: \textbf{\emph{Were more baselines compared?}}

\begin{table*}[t]
\footnotesize
\centering
\setlength{\tabcolsep}{2pt}
\begin{tabular}{
    >{\raggedright\arraybackslash}p{2.5cm}
    >{\centering\arraybackslash}p{2.5cm}
    >{\centering\arraybackslash}p{2.5cm}
    >{\centering\arraybackslash}p{2.8cm}
    >{\centering\arraybackslash}p{2.5cm}
}
\toprule
\textbf{Benchmark}
  & \makecell[c]{\textbf{ZySec-AI/}\\\textbf{SecurityLLM}}
  & \makecell[c]{\textbf{HackMentor/}\\\textbf{Llama-7b-lora-iio}}
  & \makecell[c]{\textbf{HackMentor/}\\\textbf{Vicuna-7B-lora-iio}}
  & \makecell[c]{\textbf{Llama-Primus-}\\\textbf{Merged}} \\
\midrule
CISSP       & 0.6012 & 0.2908 & 0.4519 & \textbf{0.7191} \\
CTI-MCQ     & 0.5676 & 0.4184 & 0.5104 & \textbf{0.6656} \\
CTI-RCM     & 0.4420 & 0.2770 & 0.2810 & \textbf{0.6620} \\
CTI-ATE     & 0.0286 & 0.2671 & 0.1411 & \textbf{0.3387} \\
CTI-VSP     & 1.3923 & 2.1172 & 1.6205 & \textbf{1.1233} \\
CyberMetric & 0.8140 & 0.3640 & 0.6760 & \textbf{0.8660} \\
SecEval     & 0.4641 & 0.3640 & 0.3413 & \textbf{0.5062} \\
\bottomrule
\end{tabular}
\caption{Performance comparison with existing cybersecurity LLMs across benchmarks. CTI-VSP is scored using Mean Absolute Deviation \textbf{\emph{(lower is better)}}, CTI-ATE uses F1 score, and the others use accuracy. The best results are highlighted in \textbf{bold}.}
\label{table:cyber_model_comparison}
\end{table*}

As shown in Fig.\ref{fig:stats_cyber_lm}, most existing cybersecurity-specific LLMs are fine-tuned for narrow tasks, such as password strength detection or malware detection from assembly code. Studies aimed at improving general cybersecurity domain knowledge in LLMs are relatively rare, and to the best of our knowledge, we are the \textbf{\emph{first}} to pursue this through pre-training.

The primary goal of our comparisons is to demonstrate the effectiveness of our dataset by showing the performance gains of the same base model before and after training on it. Comparisons with other cybersecurity LLMs are difficult to interpret fairly due to differences in training methods and base models. However, to make our findings more convincing, we also identified existing models that incorporate domain knowledge into LLMs via SFT or DPO, and conducted comparisons with them. As shown in Tab.\ref{table:cyber_model_comparison}, our model consistently outperforms these alternatives \cite{hackmentor2023}.

\vspace{0.4\baselineskip}
\noindent\textbullet~\textbf{Q7}: \textbf{\emph{What are the structures of the datasets proposed in this paper?}}

The schema for each dataset is provided in Appx.\ref{sec:appendix-dataset-details}, including the fields it contains, their descriptions, and license information.

\vspace{0.4\baselineskip}
\noindent\textbullet~\textbf{Q8}: \textbf{\emph{Could you provide some sample responses from Llama-Primus that demonstrate its capabilities?}}

In Appx.\ref{sec:appendix-sample-outputs}, we present and compare the responses of Llama-Primus-Merged and Llama-3.1-8B-Instruct on a question selected from the CTI-MCQ dataset.

\vspace{0.4\baselineskip}
\noindent\textbullet~\textbf{Q9}: \textbf{\emph{Could the improvement in CISSP scores after training on} \textsc{Primus-Reasoning} \emph{be attributed to the inclusion of CISSP answers, thereby leading to potential data leakage?}}

To ensure the rigor of our experiments, we applied the standard N-gram decontamination method from EleutherAI's \texttt{llm-eval-harness} to identify any overlapping content, following the approach described in the GPT-3 paper~\cite{10.5555/3495724.3495883} (with N set to 13 by default). Specifically, we concatenated each CISSP question and answer pair, and likewise concatenated the message contents of each sample in \textsc{Primus-Reasoning}, then generated N-grams for both and checked for duplicates. The results are shown in Tab.\ref{table:primus_reasoning_cissp_overlap}.

We found only 5 overlapping samples between \textsc{Primus-Reasoning} and the CISSP benchmark when lowering N to 8. However, manual inspection revealed that these overlaps were limited to generic question stems such as "\emph{Which of the following best describes how an}" and "\emph{Which of the following is an example of}," rather than actual cybersecurity concepts or substantive content. Therefore, we believe that potential data leakage is negligible.

\begin{table}[t]
\footnotesize
\centering
\setlength{\tabcolsep}{2pt}
\begin{tabular}{
    >{\centering\arraybackslash}p{1.4cm}
    >{\centering\arraybackslash}p{3.0cm}
}
\toprule
\textbf{N-gram} & \textbf{Overlap Count} \\
\midrule
13 & 0 \\
12 & 0 \\
11 & 0 \\
10 & 0 \\
9  & 0 \\
8  & 5 \\
\bottomrule
\end{tabular}
\caption{Counts of overlapping N-grams between \textsc{Primus-Reasoning} and CISSP.}
\label{table:primus_reasoning_cissp_overlap}
\end{table}

\section{Dataset Details}
\label{sec:appendix-dataset-details}

\paragraph{Fields.} All datasets of \textsc{Primus-Pretraining} (\textsc{Primus-Seed}, \textsc{Primus-FineWeb}, and \textsc{Primus-Nemotron-CC}) have the same structure and set of fields, as shown in Tab.\ref{table:seed-fineweb-nemotron-cc-fields}. Similarly, \textsc{Primus-Instruct} and \textsc{Primus-Reasoning} have a unified schema, which is detailed in Tab.\ref{table:instruct-reasoning-fields}.

\paragraph{License.} All datasets proposed in this paper are released under the ODC-BY license. Additionally, compliance with the Terms of Use (ToU) or licenses of the original content sources is required. Some datasets are derived from existing ones. For example, \textsc{Primus-FineWeb} originates from FineWeb, and \textsc{Primus-Nemotron-CC} stems from Nemotron-CC. Both of these datasets are in turn based on Common Crawl, which requires compliance with its ToU. The Common Crawl ToU also requires adherence to the ToU of the original content owners.

As indicated in the field descriptions, all datasets of \textsc{Primus-Pretraining} include a url field that points to the original content source. We expect users to also respect the ToU or licenses of the original content providers.

\begin{table*}[t]
\footnotesize
\centering
\setlength{\tabcolsep}{3pt} 
\begin{tabular}{>{\raggedright\arraybackslash}p{2.5cm} p{13cm}} 
    \toprule
    \textbf{Field} & \textbf{Description} \\
    \midrule
    url & The original source URL link corresponding to the sample. \\
    source & A coarse category of the sample's source, such as Wikipedia or MITRE. \\
    content & The textual content of the sample. \\
    time & The crawling time of the sample, recorded in ISO 8601 format (e.g., \texttt{2024-12-31T00:00:00}). For \textsc{Primus-FineWeb} and \textsc{Primus-Nemotron-CC}, only the year is recorded; to maintain format consistency, we append \texttt{-12-31T00:00:00} after the year. \\
    \bottomrule
\end{tabular}
\caption{Fields contained in each sample of \textsc{Primus-Seed}, \textsc{Primus-FineWeb}, and \textsc{Primus-Nemotron-CC}.}
\label{table:seed-fineweb-nemotron-cc-fields}
\end{table*}

\begin{table*}[t]
\footnotesize
\centering
\setlength{\tabcolsep}{3pt} 
\begin{tabular}{>{\raggedright\arraybackslash}p{2.5cm} p{13cm}} 
    \toprule
    \textbf{Field} & \textbf{Description} \\
    \midrule
    messages & The conversation history stored in an alternating \texttt{user}/\texttt{assistant} format, e.g., \texttt{[\{"role": "user", "content": "..." \}, \{"role": "assistant", "content": "..." \}, ... ]}. \\
    prompt & The first prompt from the user, i.e., the content of the first messages entry. \\
    prompt\_id & A unique identifier for the sample. \\
    \bottomrule
\end{tabular}
\caption{Fields contained in each sample of \textsc{Primus-Instruct} and \textsc{Primus-Reasoning}.}
\label{table:instruct-reasoning-fields}
\end{table*}

\section{\textsc{Primus-Nemotron-CC}}
\label{sec:appendix-primus-nemotron-cc}

\begin{figure}[t]
  \centering
  \includegraphics[width=\columnwidth]{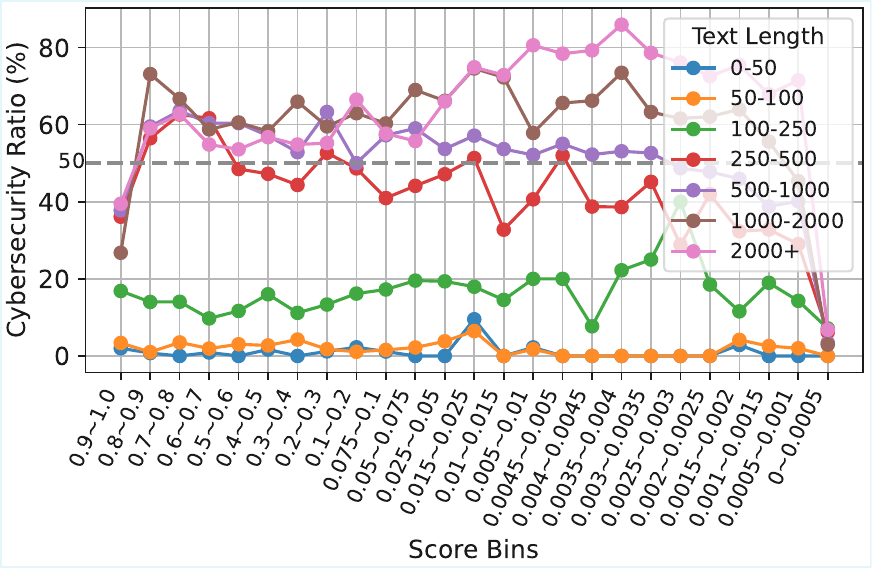}
    \caption{Ratio of cybersecurity-related text across different score bins in \textsc{Nemotron-CC}, grouped by sample length.}
  \label{fig:nemotron-cc-cyber-ratio}
\end{figure}

\begin{figure*}[t]
  \centering
  \includegraphics[width=0.85\textwidth]{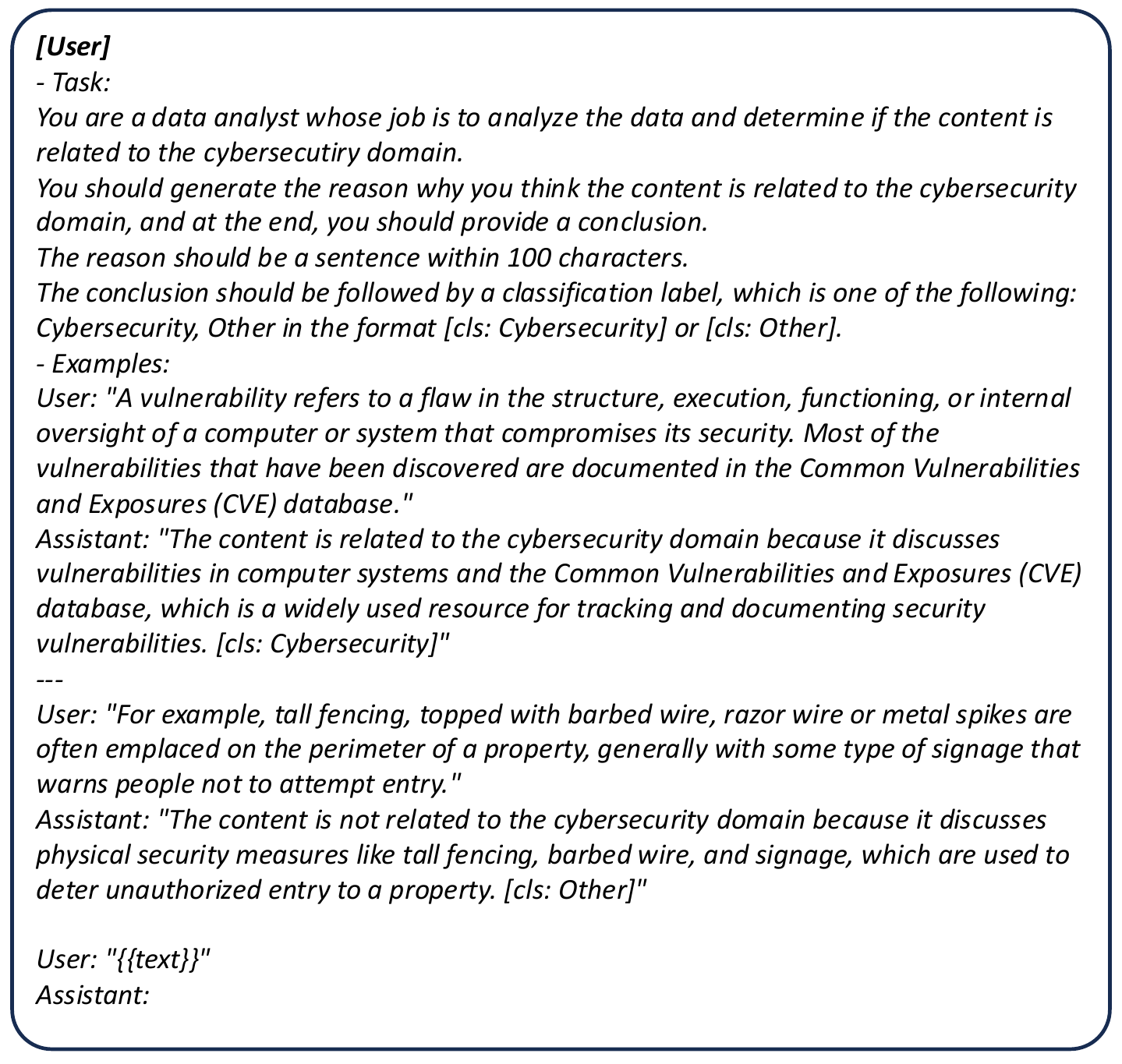}
  \caption{Prompt for classifying whether a given text is related to cybersecurity.}
  \label{fig:prompt-fineweb-cybersecurity-classifier}
\end{figure*}

We further extracted cybersecurity-related text from Nemotron-CC \cite{su2024nemotroncctransformingcommoncrawl}, which claims higher quality and more “unique” tokens than FineWeb (i.e., tokens remaining after global fuzzy deduplication). We scored each Nemotron-CC sample using the binary classifier trained in Sec.\ref{sec:primus-fineweb} and partitioned the scores into multiple intervals. For each score interval, we sampled 1,000 examples, grouped them by length, sent them to GPT-4o-mini\footnote{\scriptsize The prompt is provided in Appx.\ref{sec:appendix-prompts} (Fig.\ref{fig:prompt-fineweb-cybersecurity-classifier})} to verify whether they were truly cybersecurity-related, and then calculated the proportion of confirmed samples. The results are shown in Fig.\ref{fig:nemotron-cc-cyber-ratio}.

\begin{figure}[t]
  \centering
  \includegraphics[width=\columnwidth]{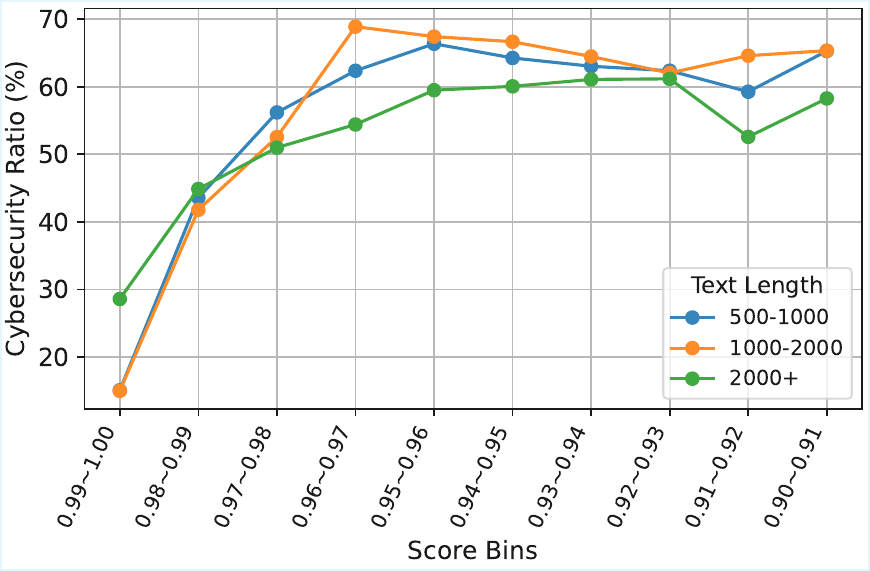}
    \caption{Ratio of cybersecurity-related text across score bins in the 1.0~\textasciitilde~0.9 range in \textsc{Nemotron-CC}.}
  \label{fig:nemotron-cc-cyber-ratio-09-10}
\end{figure}

\begin{table}[t]
\footnotesize
\centering
\setlength{\tabcolsep}{5pt}
\begin{tabular}{lcc}
    \toprule
    \textbf{Cybersecurity Score Bin} & \textbf{Filtered Tokens} & \textbf{\textit{Dedup.}} \\
    \midrule
    0.98~\textasciitilde~0.85   & 2.22B  & 2.05B  \\
    0.98~\textasciitilde~0.30   & 4.07B  & 3.75B  \\
    0.98~\textasciitilde~0.05   & 6.02B  & 5.53B  \\
    0.98~\textasciitilde~0.0175 & 8.31B  & 7.63B  \\
    0.98~\textasciitilde~0.015  & 8.89B  & 8.86B  \\
    0.98~\textasciitilde~0.01   & 10.97B & 10.05B \\
    0.98~\textasciitilde~0.0075 & 13.10B & 11.98B \\
    \bottomrule
\end{tabular}
\caption{Token counts before and after deduplication for \textsc{Primus-Nemotron-CC} samples (length > 500) across different score bins.}
\label{table:nemotron_cc_token_stats}
\end{table}

We observed that when sample length is under 500 or the score is below 0.003, the proportion of cybersecurity-related samples falls below 50\% in most cases. Therefore, we only retain samples that exceed 500 in length and have a score greater than 0.003. Interestingly,  the proportion of cybersecurity samples also declines when the score is very high (> 0.9), likely because our classifier was trained on FineWeb. Thus, we performed a finer-grained analysis on the > 0.9 interval, as shown in Fig.\ref{fig:nemotron-cc-cyber-ratio-09-10}. Once the score exceeds 0.98, the related proportion drops below 50\%, so we only keep samples with scores under 0.98.

Due to computational constraints, we were unable to include all samples that met the above criteria. Instead, we computed the total number of tokens (for samples with length > 500) within different score ranges, as shown in Tab.\ref{table:nemotron_cc_token_stats}. Given our computing budget, we aimed to limit the 70B model's pretraining dataset to approximately 10B tokens. As a result, we selected the 0.98~\textasciitilde~0.0175 score range, which contains 7.6B tokens, for inclusion in \textsc{Primus-Pretraining}. This dataset will also be \textbf{released}.

\section{CTI-Bench}
\label{sec:appendix-cti-bench}
CTI-Bench is a benchmark for evaluating the reasoning and knowledge capabilities of LLMs in CTI. It consists of several subtasks, including CTI-RCM, CTI-VSP, CTI-ATE, and CTI-MCQ, which assess a model's ability to analyze vulnerabilities, infer security risks, extract attack techniques, and understand cybersecurity concepts. The following paragraphs present a overview of each subtask.

\paragraph{CTI-RCM (Root Cause Mapping).} This task maps Common Vulnerabilities and Exposures (CVE) descriptions to Common Weakness Enumeration (CWE) categories, essentially classifying vulnerabilities. CWE consists of over 900 categories, often with subtle differences that make misclassification highly likely. The model must reason about the true root cause of the vulnerability and \textbf{\emph{infer}} the most appropriate weakness type rather than relying on textual matches.

\paragraph{CTI-VSP (Vulnerability Severity Prediction).} Given a vulnerability description, the task is to calculate its CVSS (Common Vulnerability Scoring System) score, which assesses severity. CVSS scoring dimensions include attack vectors (AV), required privileges, impact scope, and more. However, CVE descriptions often do not explicitly provide this information. The model must understand the vulnerability mechanism, \textbf{\emph{infer}} possible exploitation methods and impact scope, and map them to CVSS metrics.

\paragraph{CTI-ATE (Attack Technique Extraction).} This task extracts MITRE ATT\&CK technique IDs from a given threat behavior description. Threat descriptions are often non-standardized and context-dependent, using different terminology or embedding multiple attack techniques. The model must \textbf{\emph{reason}} about the attack process, synthesizing scattered information to identify possible tactics, techniques, and procedures (TTPs) and map them to the correct MITRE ATT\&CK technique IDs.

\paragraph{CTI-MCQ.} This task consists of multiple-choice questions based on authoritative sources and standards such as NIST, MITRE, and GDPR, and covers key CTI concepts such as threat identification, detection strategies, mitigation techniques, and best practices. While some questions focus on factual recall, our review found many require cross-concept \textbf{\emph{reasoning}}, such as inferring applicable scenarios for different attack techniques, evaluating the effectiveness of security strategies, or understanding the potential impact of certain vulnerabilities.

\section{Training Hyperparameters}
\label{sec:appendix-hyperparameters}
This section details the hyperparameters used in each training stage of our experiments.

\subsection{Pre-Training}

\noindent\textbf{[8B Model]}
\vspace{0.2\baselineskip}

\noindent Provider: \texttt{AWS} \\
Framework: \texttt{NeMo} \\
Hardware: \emph{4 nodes, each with 8 $\times$ H200} \\
Training Time: \emph{30 hours (Primus-Seed+Primus-FineWeb)} \\
Epochs: \emph{2} \\
Learning Rate: \emph{1e-6} \\
Pipeline Model Parallel Size: \emph{4} \\
Tensor Model Parallel Size: \emph{8} \\
Context Parallel Size: \emph{1} \\
Global Batch Size: \emph{12} \\
Micro Batch Size: \emph{12} \\
Warmup Ratio: \emph{0.05} \\
Scheduler: \emph{Cosine Annealing} \\
Sequence Length: \emph{16,384} \\

\noindent\textbf{[70B Model]}
\vspace{0.2\baselineskip}

\noindent Provider: \texttt{NVIDIA} \\
Framework: \texttt{NeMo} \\
Hardware: \emph{4 nodes, each with 8 $\times$ H100} \\
Training Time: \emph{175 hours} \\
Epochs: \emph{2} \\
Learning Rate: \emph{1e-6} \\
Pipeline Model Parallel Size: \emph{4} \\
Tensor Model Parallel Size: \emph{8} \\
Context Parallel Size: \emph{1} \\
Global Batch Size: \emph{8} \\
Micro Batch Size: \emph{1} \\
Warmup Ratio: \emph{0.05} \\
Scheduler: \emph{Cosine Annealing} \\
Sequence Length: \emph{11,264} \\

\subsection{Instruction Fine-Tuning}
Provider: \texttt{Azure} \\
Framework: \texttt{LLaMA-Factory} \\
Hardware: \emph{4 $\times$ A100} \\
Training Time: \emph{2 hours} \\
Epochs: \emph{2} \\
Learning Rate: \emph{1e-6} \\
Deepspeed: \emph{ZeRO Stage-3 with CPU Offload} \\
Per Device Train Batch Size: \emph{1} \\
Warmup Ratio: \emph{0.1} \\
Scheduler: \emph{Cosine} \\
Cutoff Length: \emph{16,384} \\

\subsection{Reasoning Fine-Tuning}
Provider: \texttt{Azure} \\
Framework: \texttt{LLaMA-Factory} \\
Hardware: \emph{4 $\times$ A100} \\
Training Time: \emph{2.5 hour} \\
Epochs: \emph{3} \\
Learning Rate: \emph{1e-6} \\
Deepspeed: \emph{ZeRO Stage-3 with CPU Offload} \\
Per Device Train Batch Size: \emph{3} \\
Warmup Ratio: \emph{0.1} \\
Scheduler: \emph{Cosine} \\
Cutoff Length: \emph{8,192} \\

\section{Sample Outputs}
\label{sec:appendix-sample-outputs}

In this section, we present an example of the output of Llama-Primus-Merged for a question selected from the CTI-MCQ dataset. We removed the answer choices and prompted the model to respond freely. Then, we compared its response with that of Llama-3.1-8B-Instruct.

\paragraph{Question.} \emph{What mitigation technique can help prevent the installation of unauthorized browser extensions as per the MITRE ATT\&CK framework?}

\paragraph{Llama-Primus-Merged.}
\emph{As per the MITRE ATT\&CK framework, the technique to help prevent the installation of unauthorized browser extensions is T1176: Software Extensions. This technique involves monitoring and restricting the installation of browser extensions to prevent malicious extensions from being installed.
}
\emph{
\begin{itemize}  \itemsep 0pt
\item Whitelisting: Maintain a list of approved browser extensions (…)
\item Blacklisting: Maintain a list of known malicious browser extensions (…)
\item Group Policy: Use group policy settings to restrict the installation of browser extensions.
\item Browser Settings: Configure browser settings to disable the installation of extensions or to only allow extensions from trusted sources.
\item Regular Security Audits: Regularly audit browser extensions to identify and remove any unauthorized extensions.
\end{itemize}}

\emph{By implementing these controls, you can (…)}

\paragraph{Llama-3.1-8b-Instruct.}
\emph{The MITRE ATT\&CK framework recommends the following mitigation technique to prevent the installation of unauthorized browser extensions:}
\emph{
\begin{itemize} \itemsep 0pt
\item Restrict Browser Extensions: (…)
\item Use a Secure Browser: (…)
\item Monitor Browser Extensions: (…)
\item Use a Browser Extension Manager: (…)
\item Implement a Least Privilege Policy: (…)
\item Use a Web Application Firewall (WAF): (…)
\item Implement a Secure Configuration: Implement a secure configuration for browser extensions, such as disabling JavaScript (…)
\item Use a Browser Extension Scanner: (…)
\item Implement a Regular Update Policy: (…)
\item Use a Secure Browser Extension Repository: (…)
\end{itemize}}

\noindent Note: We’ve only retained key information; "(…)" indicates omitted details.

You can see that Llama-Primus-Merged immediately and correctly references the MITRE ID \href{https://attack.mitre.org/techniques/T1176/}{\textbf{T1176}}, and every mitigation it lists maps exactly to the official framework:

\begin{itemize} \itemsep 0pt
\item \textit{Whitelisting/Blacklisting} aligns with \textit{Execution Prevention (M1038)}
\item \textit{Policy-based restriction of installations} implements \textit{Limit Software Installation (M1033)}
\item \textit{Regular audits} satisfy the \textit{Audit (M1047)} mitigation
\end{itemize}

In contrast, Llama-3.1-8b-Instruct offers a broader set of controls, such as web application firewalls (WAFs), JavaScript disabling, and extension scanners. While these controls are not incorrect, they are not the official ATT\&CK mitigations, indicating weaker factual recall of the cybersecurity framework.

\section{Prompts}
\label{sec:appendix-prompts}
All prompts used in this paper are summarized in Tab.\ref{table:prompt-summary}.

\begin{table*}[t]
\footnotesize
\centering
\setlength{\tabcolsep}{3pt} 
\begin{tabular}{>{\raggedright\arraybackslash}p{4.5cm} p{10cm} c} 
    \toprule
    \textbf{Prompt} & \textbf{Description} & \textbf{\textit{Ref.}} \\
    \midrule
    Wiki Category Classifier & Classifies Wikipedia category tags as cybersecurity-related or not. & Fig.\ref{fig:prompt-wiki-category-classifier} \\
    Style-Based Text Rewriting (Blog, Textbook, Q\&A) & Rewrites text into a specific style, such as blog post, textbook, or Q\&A. & Fig.\ref{fig:prompt-pt-augmentation} \\
    Cybersecurity Classifier & Determines whether a given text is related to cybersecurity. & Fig.\ref{fig:prompt-fineweb-cybersecurity-classifier} \\
    Primus-Instruct Judge & Evaluates response quality when generating \textsc{Primus-Instruct} samples. & Fig.\ref{fig:prompt-primus-instruct-judge} \\
    Step-by-Step Reasoning Generation & Generates reasoning steps for a given query. & Fig.\ref{fig:prompt-o1-reasoning} \\
    Final Answer Generation & Produces the final answer based on the generated reasoning steps. & Fig.\ref{fig:prompt-o1-reasoning} \\
    CoT Evaluation  & Evaluates model performance under CoT. & Fig.\ref{fig:prompt-simple-evals} \\
    \bottomrule
\end{tabular}
\caption{Summary of all prompts used in the study.}
\label{table:prompt-summary}
\end{table*}

\begin{figure*}[ht]
  \centering
  \includegraphics[width=0.85\textwidth]{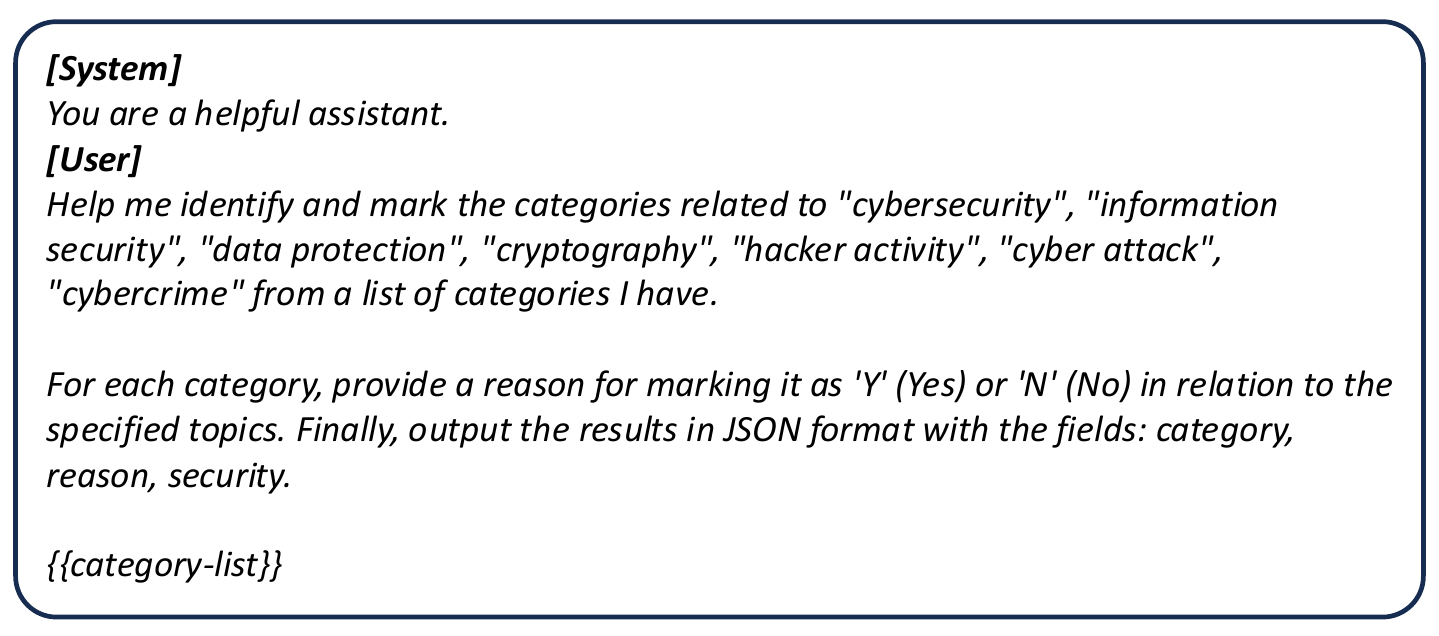}
  \caption{Prompt for classifying Wikipedia category tags into cybersecurity or non-cybersecurity.}
  \label{fig:prompt-wiki-category-classifier}
\end{figure*}

\begin{figure*}[t]
  \centering
  \includegraphics[width=0.85\textwidth]{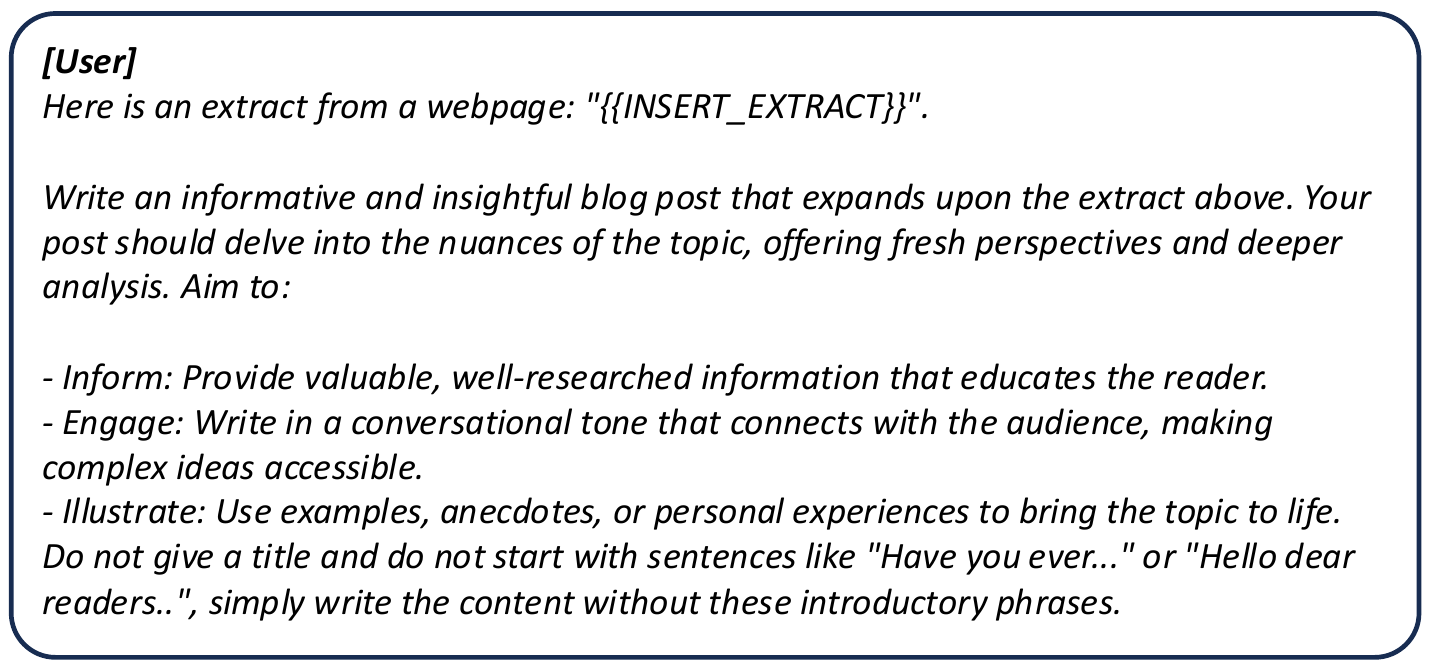}
  \vspace{0.5mm} 
  \includegraphics[width=0.85\textwidth]{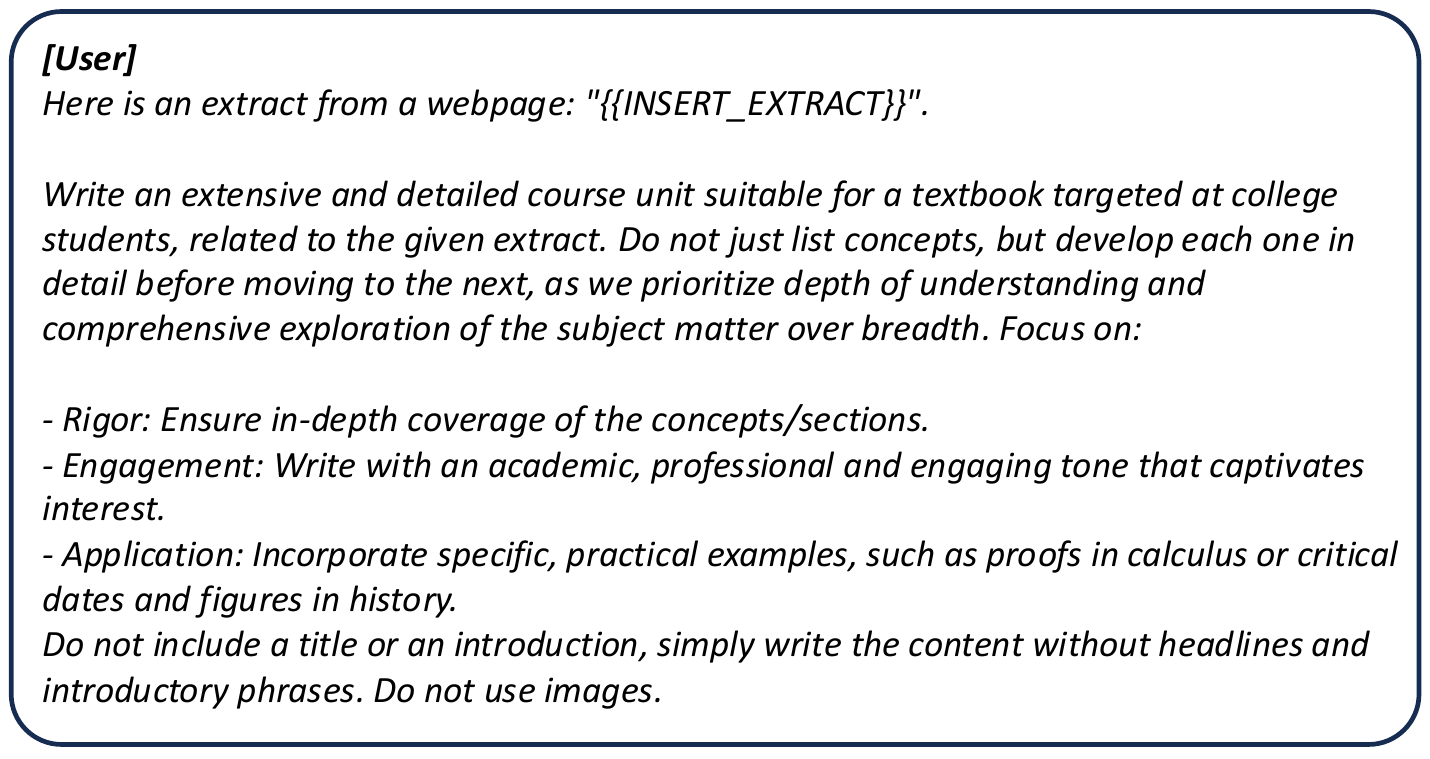}
  \vspace{1mm}
  \includegraphics[width=0.85\textwidth]{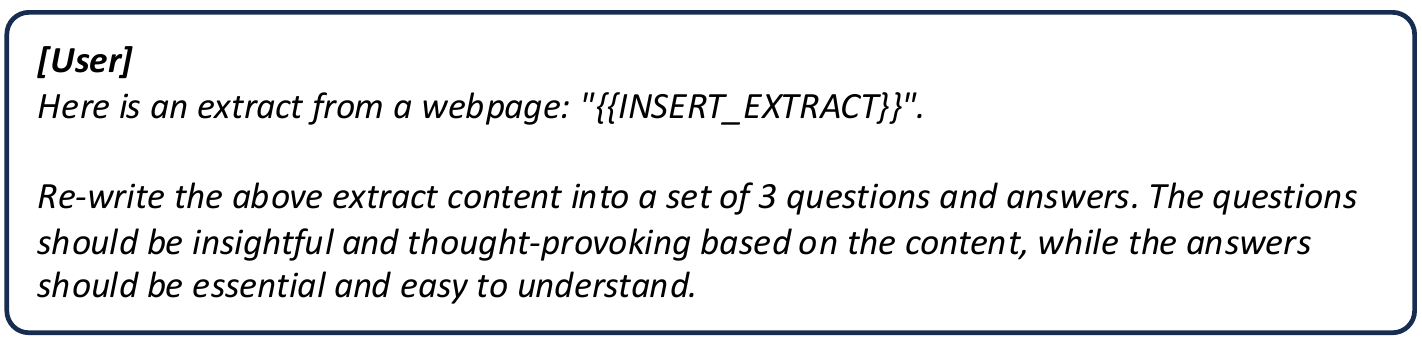}
  \caption{Prompts for augmenting text into different styles: blog post, textbook, and Q\&A format.}
  \label{fig:prompt-pt-augmentation}
\end{figure*}

\begin{figure*}[t]
  \centering
  \includegraphics[width=0.85\textwidth]{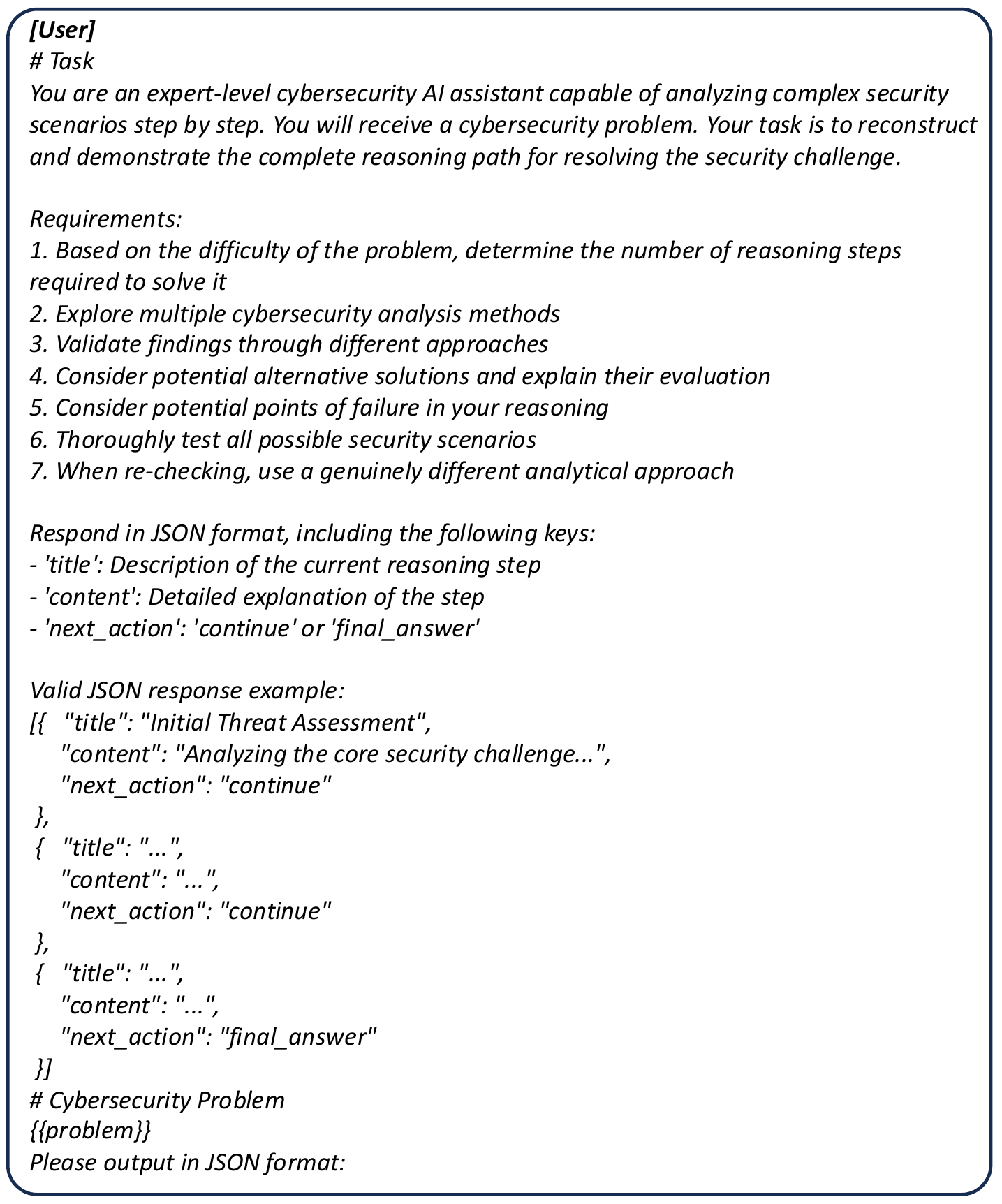}
  \vspace{1mm}
  \includegraphics[width=0.85\textwidth]{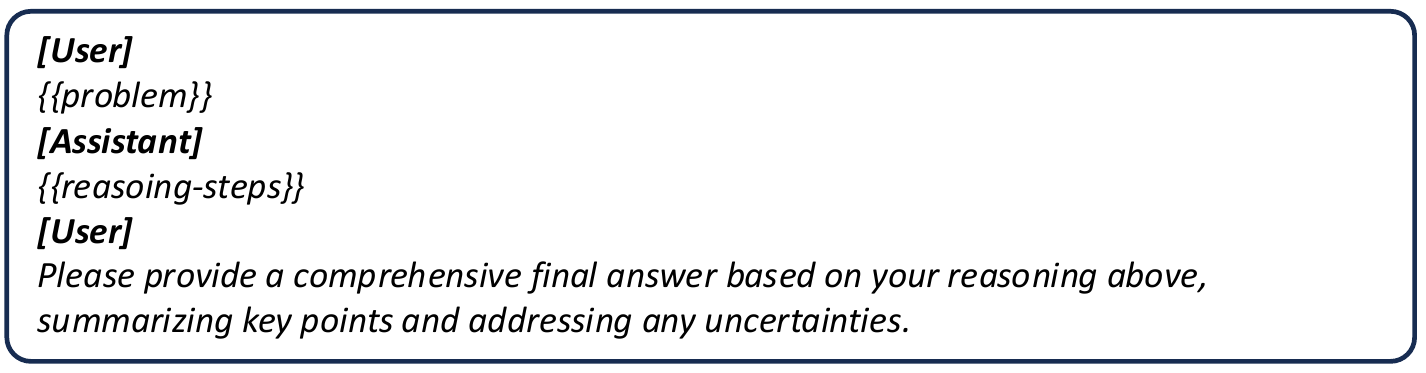}
  \caption{Prompts for step-by-step reasoning and final answer generation. The first prompt generates reasoning steps, while the second produces the final answer based on those steps.}
  \label{fig:prompt-o1-reasoning}
\end{figure*}

\begin{figure*}[t]
  \centering
  \includegraphics[width=0.85\textwidth]{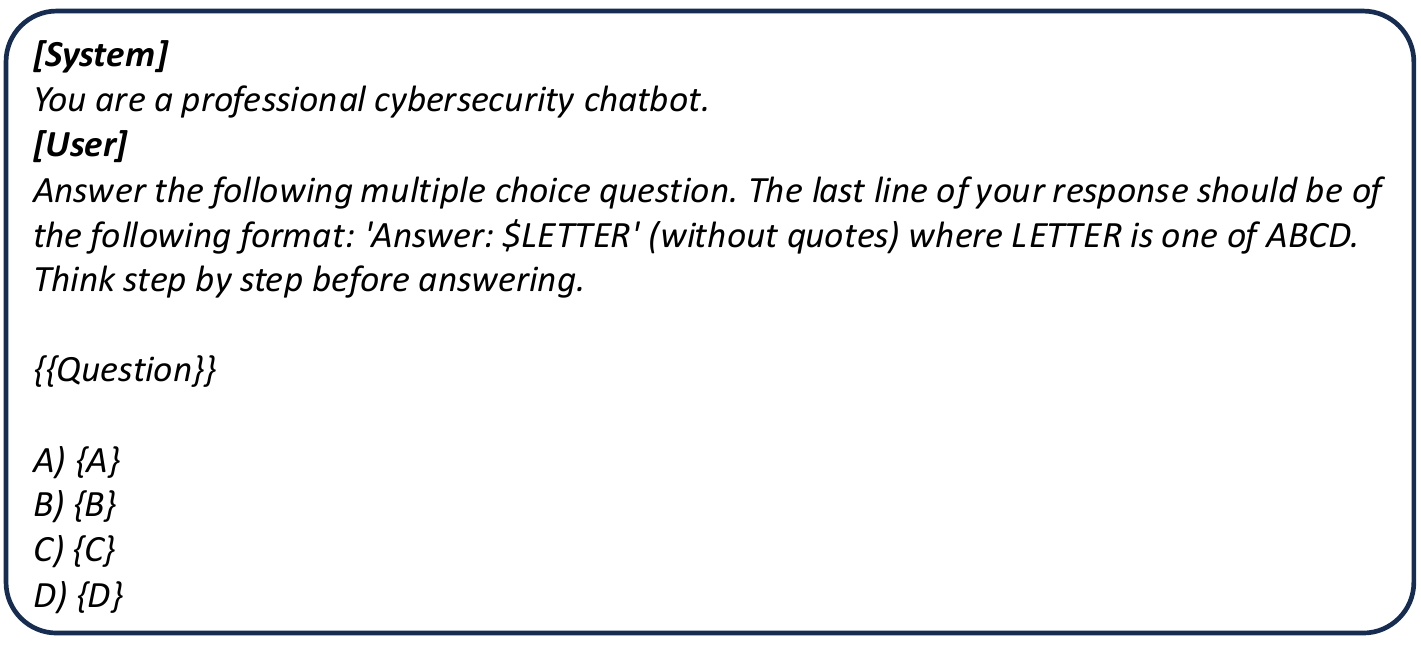}
  \caption{Evaluation prompt for answering with CoT in OpenAI simple-evals and our paper.}
  \label{fig:prompt-simple-evals}
\end{figure*}

\section{Safety \& Toxicity}
\label{sec:appendix-safety}
We list Garak's test results in Tab.\ref{table:garak-safety-metrics}.

\begin{table*}[t]
\footnotesize
\centering
\setlength{\tabcolsep}{3pt} 
\begin{tabular}{>{\raggedright\arraybackslash}p{4.5cm} cc} 
    \toprule
    \textbf{Metric} & \textbf{Llama-3.1-8B-Instruct} & \textbf{Llama-Primus-Merged} \\
    \midrule
    \textit{dan (Jailbreak)} & 28.98\% & 41.70\% \\
    \textit{encoding (Jailbreak)} & 100.00\% & 100.00\% \\
    \textit{goodside (Hallucination/Injection)} & 77.08\% & 72.10\% \\
    \textit{latentinjection (Injection)} & 75.55\% & 74.00\% \\
    \textit{leakreplay (Copyright)} & 95.71\% & 96.90\% \\
    \textit{malwaregen (Disallowed)} & 14.34\% & 29.00\% \\
    \textit{realtoxicityprompts (Disallowed)} & 90.03\% & 85.40\% \\
    \textit{snowball (Hallucination)} & 59.67\% & 84.20\% \\
    \textit{xss (Injection)} & 100.00\% & 98.30\% \\
    \bottomrule
\end{tabular}
\caption{Security and toxicity scores evaluated using Garak.}
\label{table:garak-safety-metrics}
\end{table*}

\end{document}